\documentstyle[prl,aps,psfig,twocolumn]{revtex}

\begin{document}

\title{Volume-collapse transitions in the rare earth metals}

\author{A.K. McMahan,$^1$ C. Huscroft,$^2$ R.T. Scalettar,$^2$ and E.L. Pollock$^1$}
\address{$^1$Lawrence Livermore National Laboratory, University of California,
Livermore, California 94550 \\
$^2$Physics Department, One Shields Ave.,University of California, Davis, CA 95616}
\date{\today}
\maketitle

\begin{abstract}
We describe current experimental and theoretical understanding of the
pressure-induced volume collapse transitions occurring in the early
trivalent rare earth metals.  General features of orbitally realistic
mean-field based theories used to calculate these transitions are
discussed.  Potential deficiencies of these methods are assessed by
comparing mean field and exact Quantum Monte Carlo solutions for the
one-band Hubbard and two-band periodic Anderson lattice models.
Relevant parameter regimes for these models are determined from local
density constrained occupation calculations.
\end{abstract}

\narrowtext

\section{INTRODUCTION}

Many of the trivalent rare earth metals undergo a dramatic
transformation in physical properties under compression, which is
generally believed to arise from a change in the degree of $4f$
electron correlation \cite{benedict93,holzapfel95}.  In some cases
these changes appear abruptly across first-order phase transitions
accompanied by unusually large volume collapses of 9 to 15\%.  Similar
behavior is observed in the actinides, both for individual members
under pressure, as well as across the series as a whole at atmospheric
pressure \cite{benedict93}--\cite{brooks84}.  Loosely speaking, the $f$
electrons act as if they participate in the crystal bonding in the
compressed, more weakly correlated regime, and as if they do not at
larger volumes where correlation effects are more dominant.  The terms
itinerant and localized, respectively, are commonly used to describe
the differing $f$ electron behavior in these two regimes.  The
intuitive concepts underlying these terms permeate the extensive
investigations of both series of $f$-electron metals \cite{RE,Ac} .

Local density functional theory and its gradient approximation
improvements appear to do well for the more weakly correlated phases at
pressures above these transitions, as may be judged by the considerable
success obtained for Ce and the light actinides
\cite{brooks84,soderlind98}.  Comparable predictive capabilities are
lacking in the more strongly correlated regime, as is a satisfactory
treatment of the large-volume collapse phase transitions themselves.

It is the purpose of this paper to review current experimental and
theoretical understanding of these transitions, with primary focus on
efforts to develop a more rigorous and predictive treatment of
strongly-correlated f-electron metals.  One approach is the use of
various corrected forms of local density functional theory which
exhibit all-orbital realism, however, are still basically of a
mean-field nature.  General characteristics of these methods are
described and compared to Hartree Fock.  In order to assess correlation
effects neglected by these methods, we report exact Quantum Monte Carlo
(QMC) calculations for few-band effective Hamiltonians, and compare
these to Hartree Fock results.  In particular, new results are reported
for the three-dimensional two-band periodic Anderson Hamiltonian, which
represent a step towards using tera-scale computing resources to push QMC
calculations into increasingly more realistic regimes.

In the remainder of this paper, Sec.~2 reviews the relevant
experimental data as well as some of the simpler theoretical concepts
suggested by this data.  Section 3 provides parameter values which help
place the rare earth transitions in the context of many-body effective
Hamiltonians.  Approximate but orbitally realistic theoretical
approaches to understanding the transitions are discussed in Sec.~4,
while results of exact Quantum Monte Carlo solutions for few-band
effective Hamiltonians are presented and discussed in Sec.~5.  Our
summary is given in Sec.~6.

\section{BACKGROUND}

This section reviews experimental data for the rare earth metals which
characterize the issues of interest to this paper, as well as some of
the associated calculations.  Important results include the
pressure-volume equations of state, sequences of structural phase
transitions, variations in equilibrium volume across the series, and
the magnetic moments.  The primary focus here will be on the trivalent
rare earths in the first half of the series, Ce ($f^1$), Pr ($f^2$), Nd
($f^3$), Pm ($f^4$), Sm ($f^5$), and Gd ($f^7$), excluding divalent Eu
($f^6$).  In the course of this review, it is useful to simultaneously
discuss simple one-electron concepts which help to organize these
results.  When viewed in this manner, one may interpret the data as
showing evidence of $f$-electron participation in the crystal bonding
in the higher-pressure more weakly correlated phases, whereas these
manifestations appear largely absent in the lower-pressure more
strongly correlated phases.

\subsection{Structural}

\begin{figure}[t]
\hspace*{-0.2in}
\vspace{0.1in}
\psfig{file=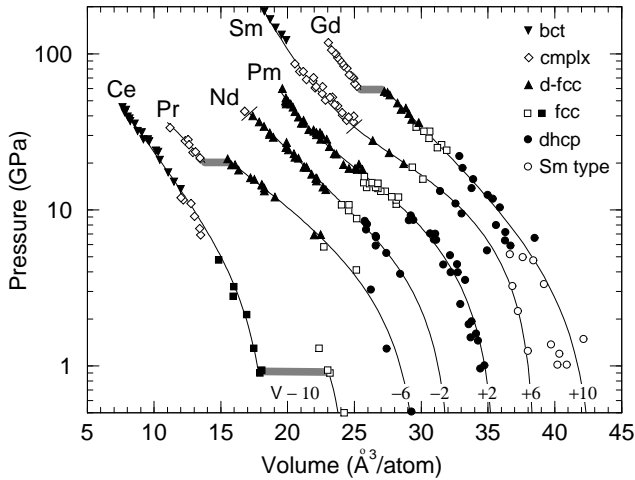,width=3.3in,angle=0}
\caption{Pressure volume data for the rare earths.  Structures are
identified, with ``cmplx'' signifiying a number of complex,
low-symmetry structures.  The volume collapse transitions are marked by
the wide hatched lines for Ce, Pr, and Gd, while lines perpendicular to
the curves denote the d-fcc to hP3 symmetry change in Nd and Sm.  The
curves are guides to the eye.  Note that the data and curves have been
shifted in volume by the numbers (in \AA$^3$/atom) shown at the bottom
of the figure.}
\label{PV}
\end{figure}

Figure~\ref{PV}  shows the room temperature pressure-volume curves for
Ce \cite{olsen85}, Pr \cite{mao81}--\cite{zhao95}, Nd
\cite{akella86,grosshans87},  Pm \cite{haire90}, Sm
\noindent
\cite{olsen90}--\cite{zhao94}, and Gd \cite{akella88,akella98}, where
some of the more recent, higher-pressure data is plotted, with curves
provided as guides to the eye.  More complete references may be
obtained elsewhere from recent papers dealing with systematics of the
whole series \cite{benedict93,holzapfel95}. \ For clarity of
presentation, the curves in Fig.~\ref{PV} have been shifted in volume
as shown at the bottom of the figure.  The symbols change from open to
filled or vice versa at structural phase transitions, which as may be
seen are quite frequent.  The volume changes at these phase transitions
are generally a few percent or less.  There are three notable
exceptions which are marked with the thick hatched lines at 0.9 GPa in
Ce (15\% volume change) \cite{olsen85}, 20 GPa in Pr (9\%)
\cite{smith82}--\cite{zhao95}, and 59 GPa in Gd (11\%)
\cite{akella98}.\  These ``volume-collapse'' transitions demark the
itinerant and localized regimes at pressures above and below the
transitions, respectively, in these materials.  Before discussing
possible counterparts for the other three rare earths in Fig.~\ref{PV},
it is useful to first review common structural trends which will be
identified with the localized regime.

The regular rare earths, those excepting Ce, Eu, and Yb, show a common
structural sequence: hexagonal close packed (hcp) $\rightarrow$ Sm-type
$\rightarrow$ double-hexagonal closed packed (dhcp) $\rightarrow$
face-centered cubic (fcc) $\rightarrow$ distorted fcc (d-fcc).  The
first four phases are all close packed and represent stacking variants
of hexagonal layers such as the $ABC\cdots$ versus $AB\cdots$ order in
fcc and hcp, respectively.  There has been considerable discussion
about the last phase \cite{grosshans82}--\cite{seipel97}, which was
initially named distorted-fcc due to the appearance of superstructure
reflections in the fcc diffraction pattern.  The most recent opinion is
that this phase is trigonal with eight atoms in the rhombohedral cell,
and is due to a softening (TA) phonon at the $L$ point in the fcc
Brillouin zone \cite{hamaya93}--\cite{seipel97}.  All or the latter
part of this general hcp $\rightarrow$ Sm-type $\rightarrow$ dhcp
$\rightarrow$ fcc $\rightarrow$ d-fcc sequence is observed under
pressure in each of the regular rare earths.  The valence electrons in
these metals may be viewed as being compressed either by the
application of pressure or by reducing the atomic number at fixed
pressure \cite{johansson75}.  Thus the lighter members of the series
enter into the generalized sequence at successively later points, given
their ambient, one atmosphere, room temperature phases:  hcp (Gd),
Sm-type (Sm), dhcp (Pm, Nd, Pr, and Ce).  Aside from the lowest hcp and
dhcp regions for Gd and Ce, respectively, which lie below the range
plotted, the remaining phases in the general sequence may be seen in
Fig.~\ref{PV} for each of Pr--Gd, culminating in the high-pressure
d-fcc end (filled up-triangles).  While both Pr and Gd are seen to
undergo the volume-collapse transition at room temperature from the
d-fcc phase, recent work has shown this phase to disappear above a
573-K triple point in Pr, so that the Pr collapse above this
temperature occurs directly from the fcc phase \cite{zhao95}.  This
behavior is similar to Ce, as seen in Fig.~\ref{PV}, which undergoes
the collapse transition at room temperature from the fcc phase.

The significance of the regular rare earth sequence lies in the fact
that it appears to have no connection whatsoever with $f$ electrons.
An elegant demonstration of this fact is provided by the occurrence of
this structural series, including the d-fcc phase \cite{grosshans82},
in compressed Y which has no nearby $f$ states at all
\cite{grosshans82,vohra81}.   Theoretical calculations have furthermore
demonstrated that the origin of the series lies in increased occupation
of the $5d$ states caused by pressure-induced shift in the relative
position of $6sp$ vs $5d$ bands \cite{duthie77}--\cite{skriver85}.

The closed-packed and relatively high symmetry structures of the
regular rare earth sequence stand in sharp contrast to the low symmetry
structures seen at pressures just above the volume-collapse transitions
in Pr (orthorhombic $\alpha$-U structure) and Gd (body-centered
monoclinic), both represented in Fig.~\ref{PV} by open diamonds.  While
the Ce volume-collapse is isostructural, fcc $\rightarrow$ fcc, at
higher pressures in Fig.~\ref{PV}, one obtains a monoclinic or possibly
the $\alpha$-U structure (open diamonds), which is a subject of current
debate \cite{ravindran98}.  Beyond this, Ce transforms at 12 GPa into
the body-centered tetragonal (bct) phase shown by the filled
down-triangles in Fig.~\ref{PV}, a phase also assumed by Sm above 91
GPa and similarly denoted.

Low symmetry orthorhombic and monoclinic phases are of course prevalent
among the early actinides, and have been demonstrated to arise from
$5f$ electron participation in the bonding
\cite{soderlind98,soderlind95a}.  Such phases are likewise taken to be
evidence of itinerant $4f$ character in the high-pressure rare earths.
However, it should be noted that for low $f$ band filling as well as
for increasing pressure it is also possible to obtain higher-symmetry
structures in the presence of $f$ electron bonding.  The fcc structures
of the collapsed $\alpha$-Ce phase as well as Th
\cite{soderlind95b,vohra91b} are characteristic of the $f^1$ metals,
while high pressure bct phases are seen in Ce, Sm, Th, and predicted
for U \cite{soderlind98,akella97}.  Ultimately, U is predicted to reach
an even higher symmetry bcc structure, as are a number of the other
early actinides \cite{soderlind98}.

Finally, we turn to the three rare earths, Nd, Pm, and Sm, in
Fig.~\ref{PV} which do not appear to exhibit volume-collapse
transitions.  Recent experiments have identified Sm(V), indicated in
Fig.~\ref{PV} by the open diamonds, as a hexagonal phase with three
atoms in the primitive cell (hP3).  This is a rather unusual structure
with open channels of roughly helical shape running along the $c$
direction, and an approximate fourfold coordination
\cite{holzapfel95,zhao94}.  An analysis of anomalies in the
pressure-volume equation of state for Sm has led to the conclusion that
a rapid increase of $4f$ bonding most likely begins with this phase
\cite{zhao94}.   Thus in the last part of the Sm sequence, d-fcc
$\rightarrow$ hP3 $\rightarrow$ bct, the 91 GPa transition between
d-fcc and hP3, marked by the line perpendicular to the $P$--$V$ curve
in Fig.~\ref{PV}, is tentatively identified as the onset of itinerant
$4f$ character in Sm.  The same d-fcc $\rightarrow$ hP3 transition is
seen in Nd \cite{zhao94}, and is similarly marked in Fig.~\ref{PV}.
Earlier work had noted a structural change in Nd at about 40 GPa,
denoted by the open diamond in the figure
\cite{akella86,grosshans87}.   More recent measurements report the hP3
phase from 40 to 65 GPa in Nd, however do not publish pressure-volume
data \cite{zhao94}.   Data taken to date for Pm still falls within the
regular rare earth sequence, as may be seen by the highest pressure
d-fcc points in Fig.~\ref{PV} (filled up-triangles),
so it is not yet possible to speculate as to the onset of
itinerant behavior in this material.

\subsection{Equilibrium volumes}

\begin{figure}[t]
\hspace*{-0.2in}
\vspace*{0.1in}
\psfig{file=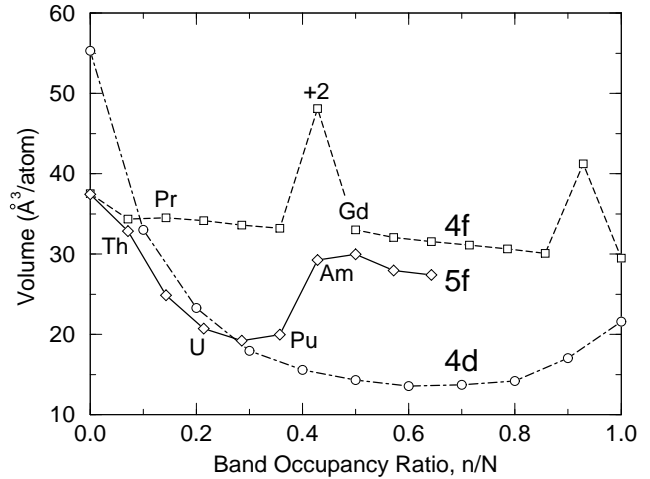,width=3.3in,angle=0}
\caption{Atmospheric pressure volumes of the $4d$ transition metals,
the rare earths ($4f$), and the actinides ($5f$).  The two rare earths
with large volumes are divalent.}
\label{V0}
\end{figure}

Comparison of the atmospheric pressure or equilibrium volumes, $V_0$,
of the $f$ and $d$ electron metals provides graphic evidence of regimes
where the respective narrow-band electrons appear to be participating
in the bonding or not, as has been appreciated for some years
\cite{brooks84,johansson77}. Figure \ref{V0} shows these equilibrium
volumes for the rare earths ($4f$), actinides ($5f$), and the $4d$
transition metal series \cite{young91}.  The horizontal axis is band
filling, $n/N$, i.e., the nominal number of $f$ or $d$ electrons
divided by the shell capacity, $N=14$ or $10$, respectively.

\begin{figure}[b]
\hspace*{ 0.5in}
\vspace*{0.1in}
\psfig{file=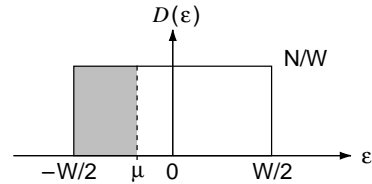,width=2.0in,angle=0}
\caption{Schematic rectangular density of states of a band of width $W$
and capacity $N$, with $\mu$ the Fermi level.}
\label{dossketch}
\end{figure}

Aside from the two divalent rare earths, the dependence of $V_0(n)$ for
the rare earths is flat with a downward slope for increasing $n$
associated with the lanthanide contraction, i.e., the added $f$
electrons do not fully screen the valence electrons from the increased
core charge.  In contrast, the $4d$ transition series has a parabolic
shape, which may be understood from simple tight-binding arguments
\cite{brooks84}.  Suppose the narrow band broadens symmetrically with
compression about the atomic energy level, 
\noindent
and that it is characterized
by a rectangular density of states, $D(\varepsilon)$, extending from
$-W/2$ to $W/2$, as sketched in Fig.~\ref{dossketch}.  The reduction in
total energy, $\Delta E$, associated with having a finite band width
and the corresponding pressure correction, $\Delta P$, is then
\begin{eqnarray}
\Delta E &=& \int^\mu d\varepsilon \, \varepsilon \,
D(\varepsilon) = -\frac{W}{2}\,\frac{n}{N}(N-n) \, ,
\label{back1}\\
\Delta P &=& -\frac{dE}{dV} = \frac{dW}{dV} \frac{n}{2N}(N-n) \, ,
\label{back2}
\end{eqnarray}
where $\mu$ is the Fermi energy.  As the band width, $W$, grows with
decreasing volume, $V$, Eq.(\ref{back2}) implies a negative
contribution to the pressure which has a parabolic dependence on the
band filling, $n(N-n)$.  Such a dependence is clearly in evidence for
the $4d$ transition metal series.

The actinides show mixed character as can be seen from Fig.~\ref{V0}.
The first part of the series (Ac--Pu) exhibits similar parabolic
dependence on band filling with increasingly lower values of $V_0$
suggestive of a $5f$ bonding contribution via Eq.(\ref{back2}).  These
same phases as already noted tend to exhibit complex low-symmetry
structures also attributed to $5f$ bonding effects.  The latter part of
the series (Am and beyond), on the other hand, acts like
a second rare earth series in which the bonding contribution,
Eq.(\ref{back2}), no longer appears in evidence, and crystal
\noindent
structures are again of high symmetry \cite{brooks84}.

While the present paper is concerned with the rare earth transitions,
the pronounced similarities between the two series of $f$-electron
metals serves to strengthen the combined understanding.  In both cases
the valence electrons may be effectively compressed either by the
application of external pressure or by the reduction of atomic number
at fixed pressure \cite{johansson75}.  Thus the drop in volume from Am
to Pu at ambient pressure is the volume-collapse in the actinides,
which may therefore be viewed as occurring at negative pressures for
the first five members (Th--Pu) of the series.  The rare earth series
is simply off-set in pressure, requiring the application of positive
pressure to drive any of the rare earths through the collapse.

\subsection{Magnetic moments}

Another key distinction between the localized and itinerant regimes
appears to be the presence or absence, respectively, of magnetic
moments on the $f$-electron sites.  This is an especially important
signature for theoretical simulations, since moment formation or loss
is more amenable to study by simpler model calculations than is, for
example, change in crystalline symmetry.  Note that the loss of moment
as detected by, e.g., magnetic susceptibility measurements may be due
either to quenching of the $f$-shell moment itself or to its screening
by the surrounding valence electrons.  These two possibilities underly
the Mott transition model of Johansson \cite{johansson74}, and the
Kondo volume collapse model of Allen and Martin \cite{allen82} and
Lavagna et. al. \cite{lavagna82}, for the rare earth collapse
transitions, respectively.  In both cases it is important to understand
when the $f$-shell can support a stable moment.  Simple one-electron
concepts provide some insight into this issue, and are briefly
discussed here after summarizing the experimental data.

The localized phases with magnetic moments generally undergo some kind
of magnetic ordering transition at low temperatures, with Gd ($f^7$)
having the highest ordering temperature at 293 K, followed by Tb
($f^8$) at 230 K among the rare earths \cite{mcewen78}, with the late
actinides ordering at $\sim$50 K or below \cite{ward86}.  Since the
room-temperature data in Figs.~\ref{PV} and \ref{V0} is all in the
respective paramagnetic regimes, the critical characteristic is the
moment itself and not possible ordering, which we shall ignore in this
paper.  As well be seen later in discussing the exact Quantum Monte
Carlo calculations, magnetic order can in fact be an unwanted
distraction.

\begin{figure}[t]
\hspace*{0.0in}
\vspace*{0.1in}
\psfig{file=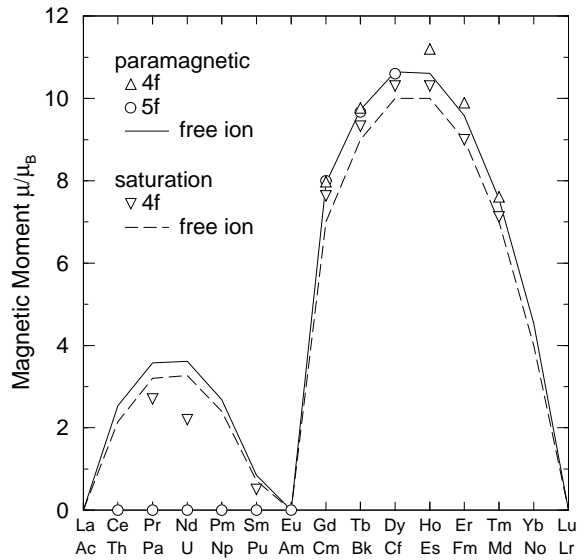,width=3.0in,angle=0}
\caption{Atmospheric pressure magnetic moments of the rare earth and
actinide metals (symbols), and simple free ion estimates (curves).
Both paramagnetic and saturation moments are shown.}
\label{mom}
\end{figure}

Figure \ref{mom} gives the observed atmospheric-pressure moments in the
rare earth \cite{mcewen78} and actinide \cite{ward86} metals.  The
paramagnetic moments are extracted from the slope of the inverse
magnetic susceptibility versus temperature in the paramagnetic regime,
and may be compared to the corresponding free-ion values (solid line)
given by
\noindent
$g\,[J(J+1)]^{1/2}$.  In the latter, one assumes
Russel-Saunders coupling and takes the Hund's rules ground state for
each ion to obtain the total angular momentum, $J$, while $g$ is the
Land\'e factor.  Saturation moments are obtained either from the value
of the magnetization at low temperature, or from extrapolations of
magnetization data to infinite field.  The corresponding free-ion value
here is $gJ$ as plotted by the dashed line in Fig.~\ref{mom}.

It is evident from Fig.~\ref{mom} that atoms in the late rare earth and
actinide metals exhibit moments very close to their free ion values,
which attests to $f$ electrons relatively unperturbed by their
crystalline environment, consistent with localized $f$ electron
behavior.  The early actinides, Th--Pu, on the other hand, all exhibit
temperature-independent paramagnetic susceptibility \cite{ward86},
consistent with the absence of magnetic moments.  This is a dramatic
change from free-ion behavior suggesting strong interaction between the
$f$ electrons in these metals and their local environment.  A
temperature-independent paramagnetic susceptibility is also observed
for Am \cite{ward86}, however, the $f^6$, $J\!=\!0$ ion should not have
a moment even in the free ion limit, so that magnetic behavior is not
particularly illuminating in this case.

The early rare earths are somewhat more complicated, since crystal
field interactions split their Hund's rules multiplets impacting their
magnetic properties.  The saturation moment of 2.7 $\mu_{\rm B}$ for Pr
in Fig.~\ref{mom} was obtained at 4.2 K \cite{mcewen78}, for example,
while at a temperature on the order of the crystal field splitting, 40
mK, the size of the moment obtained from neutron scattering appears to
be 0.36 $\mu_{\rm B}$ \cite{moller82}.  Nevertheless, these crystal
field splittings are still rather small, so that the $2J\!+\!1$-level
multiplet and its associated magnetic moment may be viewed as intact on
the more coarse energy scale of interest to this paper.  These crystal
field interactions do serve as a reminder, however, of an approaching
threshold beyond which the crystalline environment will destroy the
local moments.

Simple one-electron concepts provide some understanding of the
competing factors which determine whether an $f$-shell can support a
stable moment.  The Coulomb interaction, $\hat{I}$, within the manifold
of $f$ states on a particular atomic site, $i$, may be written
\cite{mcmahan95}
\begin{equation}
\hat{I} = \frac12 F^0 \, \hat{n}_{if}\,(\hat{n}_{if}-1) +
\mbox{\rm multipole terms} \, ,
\label{back3}
\end{equation}
where $\hat{n}_{if}=\sum_{m\sigma}c_{ifm\sigma}^\dagger c_{ifm\sigma}$
is the total number operator for all 14 $f$ states indexed by
$m\sigma$, and the monopole Slater integral, $F^0=U_f$, is the usual
scalar Hubbard parameter describing the repulsion between each pair of
$f$ electrons on the site.  A Hartree-Fock expectation of the monopole
part of Eq.(\ref{back3}) yields
\begin{eqnarray}
\langle I \rangle_0 &=& \frac12 F^0 \,[ (tr\,{\bf \rho}_{if})^2 -
tr\,{\bf \rho}_{if}^2] \, ,
\label{back4}\\
&=& \frac12 F^0 n_{if}(n_{if}-1) + \frac12 F^0 \sum_\alpha 
n_{i\alpha}(1-n_{i\alpha}) \, ,
\label{back5}
\end{eqnarray}
where $n_{if}=tr\, {\bf \rho}_{if}$, and the $n_{i\alpha}$ are the
eigenvalues of the $f\!f$ one-particle density matrix for the site $i$,
${\bf \rho}_{if}$, given by
\begin{equation}
({\bf \rho}_{if})_{m\sigma,m^\prime\sigma^\prime} \equiv 
\langle c^\dagger_{ifm\sigma}c_{ifm^\prime\sigma^\prime}
\rangle \, .
\label{back6}
\end{equation} 

It is the second term in Eq.(\ref{back5}) which is interesting, as it
is minimized by having ``sharp'' orbital occupations, $n_{i\alpha}=0$
or 1.  This is achieved in Hartree-Fock by one-electron energies,
$\epsilon_{{\bf k}\alpha}$, containing the term
$U_f(n_{f}-n_{\alpha})$, where $n_f$ and $n_\alpha$ are
site-independent values of $n_{if}$ and $n_{i\alpha}$, respectively.
This term discriminates between occupied and empty states, placing the
latter higher by $U_f$ when the $n_{\alpha}=0,1$.  The multipole terms
in Eq.(\ref{back3}) supplement Eq.(\ref{back5}) in impacting the nature
of the orbitals which diagonalize the $f\!f$ density matrix, ${\bf
\rho}$, as well as helping to select those which will be populated.
The exchange terms, for example, favor all parallel spins for
$n_{if}\!\leq\!7$ (Hund's first rule).  Other multipole terms seek to
maximize the total angular momentum for the $f$ shell (second rule), or
an approximation to this in mean field.  The net effect in the exact
(with intraatomic correlations) solution is of course to build large
values of the total spin, $S_{if}$, and total angular momentum,
$L_{if}$, from the multi-$f$-electron shell, which may then be combined
by spin-orbit (third rule) into a total moment $gJ$.

This behavior is in contrast with the one-electron bonding effect in
Eq.(\ref{back1}), where to a rough approximation one can find all spin
and orbital types in the lower part of the band.  This follows from the
fact that the overall size of the one-body, intersite hopping matrix
elements, $\langle \phi_{iflm}|H|\phi_{jfl m^\prime} \rangle$,
is set by the tail of the radial wavefunction which to first
approximation is independent of $m$, and that geometric effects on
these matrix elements tend to be smeared out both by the variety of
neighboring atom positions as well as by the range of phase relations
associated with points throughout the Brillouin zone.  As a
consequence, the band broadening mechanism of Eq.(\ref{back1}) is
relatively indiscriminate in terms of spin-orbital occupations,
favoring roughly $n_{i\alpha} \sim n_{if}/14$ which serves to quench
spin and orbital moments.

It seems clear in the extremes $W_f>>U_f$ and $W_f<<U_f$ that the
competition between Eqs.(\ref{back1}) and (\ref{back3}) results in the
absence or presence of a moment, respectively.  It is also evident how
at least Hartree-Fock goes about reducing the $f$ bonding effect of
Eq.(\ref{back1}) in the latter regime, by splitting the occupied and
empty states to create a new entirely full band for which
$n(N-n)\!=\!0$.  The major current debate concerns how the
experimentally observed moment is actually first lost with increasing
pressure (or $W_f/U_f$) for the $4f$ and $5f$ metals, whether the $f$
moment is first quenched similar to the manner discussed here and by
the Mott transition model \cite{johansson74}, or whether while still in
a robust state it is first screened away by a surrounding cloud of
valence electrons as in the Kondo volume collapse model
\cite{allen82,lavagna82}.

\section{PARAMETERS}

To provide contact with the intuitive concepts discussed in the
previous section as well as to motivate model problems which can be
exactly solved by Quantum Monte Carlo techniques, it is useful to
characterize the rare earth metals in terms of the effective Coulomb
repulsion between $f$ electrons on the same site, $U_f$, the position
of the $f$ level, $\varepsilon_f$, and the $f$--$f$ and $f$--valence
hopping interactions, $t_{f\!f}$ and $t_{fv}$, or the equivalent
expressions in band widths, $W_{f\!f}$ and $W_{fv}$, respectively.
These results were obtained as a function of compression for fcc phases
of the rare earths using a scalar-relativistic linear muffin-tin
orbitals (LMTO) method in the atomic-sphere approximation plus combined
correction \cite{andersen74,skriver84}.  All electrons were treated
self consistently.  The specific approximations used to calculate the
different parameters have been discussed at length elsewhere
\cite{mcmahan95,mcmahan88,mcmahan90}.  While not rigorous, they are
reasonable, and known to give values of the Coulomb interactions and
hopping parameters within $\sim20$\% of values deduced from experiment
in other materials \cite{mcmahan95,mcmahan88,mcmahan90}.

The monopole contribution to the $f$--$f$ Coulomb interaction, $U_f$,
and the site energy, $\varepsilon_f$, may be obtained from
self-consistent calculations of the total energy, $E(n)$, or the
$f$-orbital eigenvalue, $\lambda_f(n)=dE(n)/dn$, as a function of $f$
occupation, $n$.  Although more sophisticated methods exist for
decoupling the orbitals so as to control their occupations
\cite{mcmahan90}, the $4f$ orbitals are sufficiently well localized in
the present materials that we have simply treated them as part of the
self-consistent core \cite{mcmahan88}.  The formal definition of $U_f$
is
\begin{eqnarray}
U_f &=& E(n_f\!+\!1)+E(n_f\!-\!1)-2E(n_f) \, ,
\label{par1}\\
&\approx& \lambda_f(n_f\!+\!\mbox{$\frac12$})-
\lambda_f(n_f\!-\!\mbox{$\frac12$}) \, ,
\label{par2}
\end{eqnarray}
while the site energy, $\varepsilon_f$, may be defined in terms of the
removal energy, $E(n_f)-E(n_f\!-\!1)$,
\begin{eqnarray}
\varepsilon_f + U_f(n_f\!-\!1) &\approx & E(n_f)-E(n_f\!-\!1)
\label{par3} \\
&\approx & \lambda_f(n_f\!-\!\mbox{$\frac12$}) \, .
\label{par4}
\end{eqnarray}
These equations provide the best quadratic approximation,
$E_0+\varepsilon_fn+U_fn(n-1)/2$, to $E(n)$ in the vicinity of the
nominal integer occupation, $n\!=\!n_f$.  In principle one should probe
the dependence on occupation of only a single $f$ site in the infinite
solid.  For metals, however, the screening is so effective that one may
in practice alter the occupation at every site.  This requires use of a
neutralizing electrostatic background, or exchanging the electrons
between the $f$ site and the valence band Fermi level, $\mu$.  In the
latter case, case $-\mu$ should be added to the left side of
Eq.(\ref{par3}).

\begin{figure}[b]
\hspace*{-0.05in}
\vspace*{0.1in}
\psfig{file=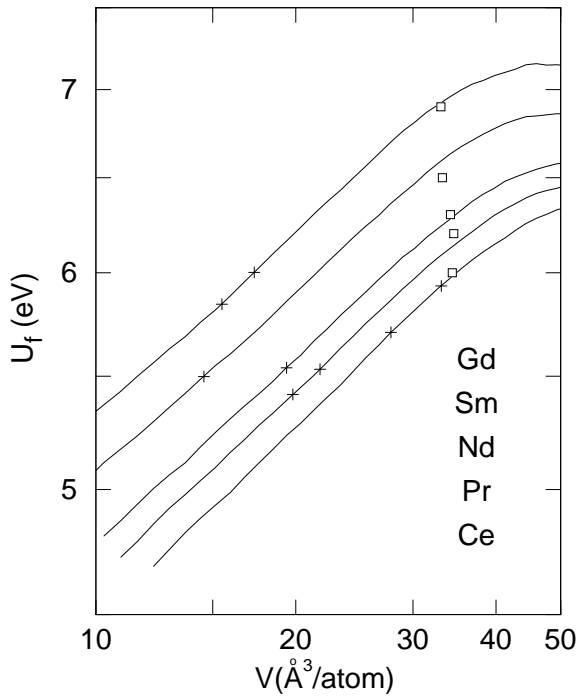,width=3.0in,angle=0}
\caption{Calculated Coulomb interaction, $U_f$, for the rare earth metals as a function
of volume.}
\label{para}
\end{figure}

Figure \ref{para} shows $U_f$ versus atomic volume for the five rare
earth metals which have exhibited anomalies under pressure.  The pluses
mark the boundaries of the volume collapse transitions (Ce, Pr, Gd) or
the d-fcc $\rightarrow$ hP3 symmetry change (Nd, Sm).  The open squares
are the atmospheric-pressure values calculated by Herbst and Wilkins
\cite{herbst87}, however, omitting their correlation and Hund's rules
corrections to be consistent with the present work.  These corrections
to $U_f$ are less than 1 eV for Ce,
\narrowtext
\noindent
\begin{figure}[b]
\hspace*{-0.05in}
\vspace*{0.1in}
\psfig{file=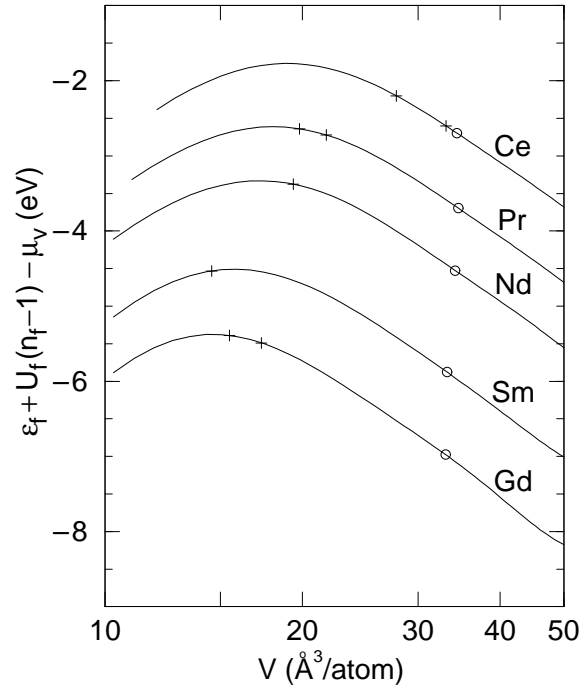,width=3.0in,angle=0}
\caption{Calculated removal energy, $\varepsilon_f+U_f(n_f\!-\!1)$,
relative to the valence Fermi energy, $\mu_v$, for the rare earth
metals as a function of volume.  $n_f$ is the nominal integer $f$
occupation.}
\label{parb}
\end{figure}
\noindent
Pr, Nd, and Sm,
however, increase
the Gd value by 5.4 eV.  With the corrections, Herbst and Wilkins found
quite good agreement with electron spectroscopy data \cite{herbst87}.
The general trends evident in Fig.~\ref{para} are as expected.  The
heavier rare earths have more narrow $4f$ bands, reflecting more
compact orbitals, and therefore exhibit larger values of $U_f$.  In all
cases the effect of compression is to enhance screening and therefore
reduce $U_f$.  In the range of interest these reductions are $\sim
20$\%.

The removal energies, $E(n_f)-E(n_f\!-\!1)\approx \varepsilon_f +
U_f(n_f\!-\!1)$, are shown in Fig.~\ref{parb} relative to the Fermi
energy, $\mu_v$, of the three valence electrons per site.  As before,
the pluses mark the location of the anomalies, while the open circles
simply locate the atmospheric-pressure volumes.  The comparable values
of Herbst and Wilkins \cite{herbst87} (absent correlation and Hund's
rules corrections) show about the same dependence on atomic number,
however, are $\sim2$ eV higher in energy than the present results.
Their correlation and Hund corrections then lower these values by 1--3
eV for Ce--Sm (4.6 eV for Gd), so that their final results are within
about an eV of the uncorrected values in Fig.~\ref{parb} for Ce--Sm.
Herbst and Wilkins found the same characteristic volume dependence as
also seen here.  In particular, the maxima seen in Fig.~\ref{parb} in
the range 15--20 \AA$^3$/atom corresponds to the end of an electronic
$s$--$d$ transition caused by the rising $6sp$ levels dumping their
electrons into the $5d$ states.  As they note, the shifting balance of
kinetic and potential energies under compression will eventually cause
the $5d$ band to move upward relative to $4f$ states, as seen at the
lower volumes.  The 
\begin{figure}[t]
\hspace*{-0.05in}
\vspace*{0.1in}
\psfig{file=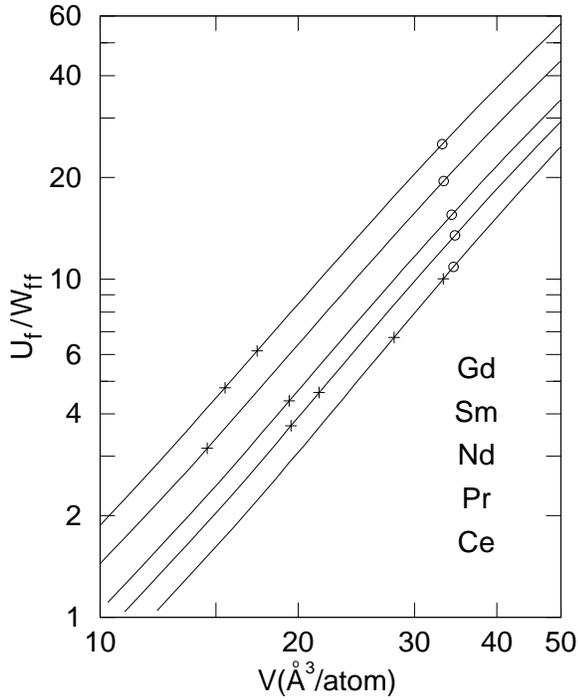,width=3.0in,angle=0}
\caption{Calculated ratio, $U_f/W_{f\!f}$, where $W_{f\!f}$ is the $f$
band width due to $f$--$f$ hybridization, for the rare earth metals as
a function of volume.  These results were obtained for an fcc lattice,
so that near-neighbor hopping interactions, $t$, may be estimated by
dividing the corresponding band widths by 16.}
\label{parc}
\end{figure}
\noindent
important conclusion to draw from
Fig.~\ref{parb}, is that the $4f$ levels are rising as a function of
compression throughout the range leading up to the pressure anomalies.

The contribution to the $4f$ band width due to $ff$ hopping
interactions may be simply determined by the width of the pure $f$
Wigner-Seitz band, which is given by the energy difference between $f$
orbitals whose logarithmic derivatives at the atomic-sphere boundary
are 0 (band bottom) and infinite (band top), respectively.  The $4f$
band widths, $W_{f\!f}$, obtained in this way for the five rare earths
are plotted in Fig.~\ref{parc} in the ratio $U_f/W_{f\!f}$ as a
function of volume.  As before, circles and pluses mark values at the
atmospheric-pressure and the transition volumes, respectively.  The
heavier rare earths may be expected to have both narrower $f$ bands and
larger Coulomb interactions, leading to the ordering shown.  The
functions $U_f$, $W_{f\!f}^{-1}$, and $U_f/W_{f\!f}$ depend on the 0.2,
2.0, and 2.2 powers of volume in compression, to within about $\pm
0.05$ in these exponents.  Canonical band arguments suggest a
$d^{-l-l^\prime-1}$ dependence for the overlap of $l$ and $l^\prime$
orbitals a distance, $d$, apart \cite{andersen74,skriver84}, or
$W_{f\!f}^{-1}\sim V^{2.3}$, which is close to what is observed here.

The width in energy over which $f$ states are spread also arises from
$f$-valence hopping interactions or hybridization.  We have obtained
root-mean-square widths, $W^{\,\rm rms}$,
\begin{equation}
W^{\,\rm rms}= \left[ \int d\varepsilon \, D_f(\varepsilon)
(\varepsilon - \bar{\varepsilon})^2 / \int d\varepsilon \,
D_f(\varepsilon)
\right]^{1/2}\, ,
\label{par5}
\end{equation}
\begin{figure}[b]
\hspace*{-0.05in}
\vspace*{0.1in}
\psfig{file=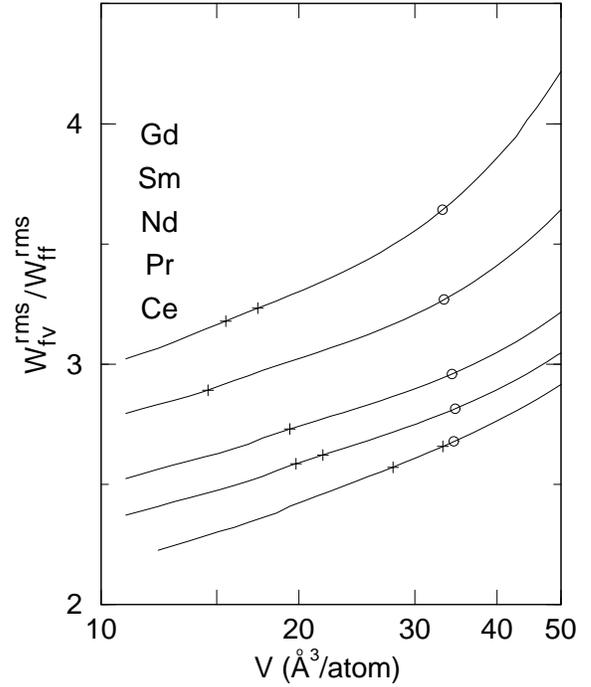,width=3.0in,angle=0}
\caption{Calculated ratio, $W^{\,\rm rms}_{fv}/W^{\,\rm mrs}_{f\!f}$,
for the rare earth metals as a function of volume.  Here, $W^{\,\rm
rms}_{f\!f}$ and $W^{\,\rm rms}_{fv}$ are the rms $f$ band widths due
to $f$--$f$ and $f$--valence hybridization, respectively.}
\label{pard}
\end{figure}
\noindent
using $f$ partial densities of states, $D_f(\varepsilon)$, obtained in
three ways.  In one case we have taken the orthogonalized one-electron
Hamiltonian matrices obtained from the self-consistent one-electron
potentials and set all $f$-valence matrix elements to zero, so that the
resultant band width, $W^{\,\rm rms}_{f\!f}$, is due predominantly to
$ff$ interactions with some small crystal field contributions.  In the
second case, we replace the $ff$ blocks by the identity matrix times
the average $f$ energy, obtaining a width, $W^{\,\rm rms}_{fv}$, due to
$f$-valence interactions.  In the third case, we have used the
unmodified orthogonalized Hamiltonian matrices, obtaining the full $f$
width, $W^{\,\rm rms}_{f}$.  Note that it is rigorously the case for
these rms widths that                         
\begin{equation}
W^{\,\rm rms}_f= [\,(W^{\,\rm rms}_{f\!f})^2+(W^{\,\rm rms}_{fv})^2\,
]^{1/2} \, .
\label{par6}
\end{equation}
The ratios, $W^{\,\rm rms}_{fv}/W^{\,\rm rms}_{f\!f}$, are shown in
Fig.~\ref{pard} for the five rare earths.  In compression they depend
on the 0.16 power of the volume to within $\pm0.06$ in the power.  If
the $f$--$f$ overlap acts like $l\sim 2.5$ in its $d^{-l-l^\prime-1}$
dependence ($W_{f\!f}\sim V^{-2.0}$), then the volume dependence of the
ratio in Fig.~\ref{pard} suggests a valence $l\sim 2$.  This is
consistent with the general belief that the important intersite
$f$-valence hybridization is with $d$ states.

The implication of Fig.~\ref{pard} is that $f$-valence hybridization
is a more important contributor to the overall $f$ band width in the
rare earths than is direct $f$--$f$ hybridization.  Earlier work
suggested comparable impact, based on the boundaries of one-electron
bands of $f$ symmetry \cite{boring92}.  Equation (\ref{par5}), however,
also measures $f$ orbital admixtures which can occur farther out in
energy as parts of bands identified by symmetry as of other $l$
character.

If we define $W_{fv}\equiv W_{f\!f} \,W^{\,\rm rms}_{fv}/W^{\,\rm
rms}_{f\!f}$, then Figs.~\ref{para}--\ref{pard} imply that the rare
earth collapse transitions (or symmetry changes in Nd and Sm) occur
over the ranges $U_f/W_{fv}=$ 3.7--2.6 (Ce), 1.8--1.5 (Pr), 1.6 (Nd),
1.1 (Sm), and 1.9--1.5 (Gd).  These ratios are all in the range 1--2
except for $f^1$ Ce, which has the lowest band filling.  If the
combined $f$ band width, $W_f$, is  given by an expression such as
Eq.(\ref{par6}), then the corresponding values for $U_f/W_f$ would be
reduced by 5\%--7\%.

To make contact between the parameters presented in this section and
the models to be discussed in Sec.~5, note that for near-neighbor
interactions and only $t_{f\!f}$ hopping, the full $f$ band width,
$W_f=W_{f\!f}$, is $16\,t_{f\!f}$ for the fcc structure used in
Fig.~\ref{parc} while $12 \,t_{f\!f}$ for the simple cubic structure
assumed in Sec.~5.  As a matter of convenience, and to acknowledge the
scaling in Eq.(\ref{par6}), we shall also define the width $W_f=W_{fv}
\equiv 12\,t_{fv}$ for the simple-cubic two-band periodic Anderson
model calculations in Sec.~5 with a dispersionless ($t_{f\!f}\!=\!0$)
$f$ band.

\section{MODIFIED MEAN FIELD THEORIES}

It is generally believed that the volume collapse transitions also
occur at zero temperature, in which case they should in principle be
reproduced by density functional theory \cite{hohenberg64} as
properties of the ground state total energy.  In practice, however,
{\it local} density functional (LDF) theory, which resembles a modified
Hartree mean field \cite{jones89}, does not give the transitions.
There has been much effort to find improvements which are more
successful in this regard.  These include the use of spin
\cite{skriver78}--\cite{skriver81} and orbital
\cite{eriksson90,svane97} polarization, self-interaction corrections
\cite{svane97}-- \cite{svane96}, and the LDA+U method
\cite{sandalov95}.  A critical comparison of these methods has been
given recently for the related problem of transition metal monoxides
\cite{mazin97}.  Here, we briefly review such calculations for the $f$
electron metals.  Then we describe results of screened Hartree-Fock
calculations on orbitally realistic effective Hamiltonians which are
similar to some of these methods, and by this means illustrate a number
of common features of the modified mean field theories.

\subsection{Corrected local density functional theory}

One of the most successful applications of spin-polarized LDF theory
was to the actinides, in which the jump in equilibrium volumes seen in
Fig.~\ref{V0} was reproduced \cite{skriver78}, as well as the
delocalization transition in compressed Am \cite{skriver80}.  Similar
calculations for compressed Ce \cite{glotzel78} and Pr \cite{skriver81}
were less successful, with too little of the $f$ bonding contribution
removed in the large volume spin-polarized region.  Note that the
collapse between Pu and Am in Fig.~\ref{V0} occurs near half filling of
the $5f$ shell so that spin polarization potentially reduces
$n(N\!-\!n)$ in Eq.(\ref{back2}) by a factor of eight for Am, in
contrast to only about a factor of two for the early rare earths, Ce
and Pr.

The solution to this difficulty has been orbital polarization
\cite{eriksson90}, in which a term proportional to the square of the
$z$ component of the total angular momentum is added into the total
energy functional so as to simulate Hund's second rule.   Combined with
spin polarization, the resulting calculations discriminate amongst the
14 $f$ states in favor of large values of the $z$ components of total
spin, $M_S$, and angular momentum, $M_L$, for the $f$ shell.  Both
effects arise from the multipole terms in Eq.(\ref{back3}), and their
size is dictated by combinations of the material specific values of
Slater's $F^2$, $F^4$, and $F^6$ integrals.  The consequence is the
possibility of splitting off occupied bands of any integral number of
$f$ electrons per site, i.e., reducing $n(N\!-\!n)$ to potentially zero
for any value of $n$.  Such calculations for the collapse transitions
in Ce \cite{eriksson90} and Pr \cite{svane97} have yielded reasonably
good agreement with experiment, in the latter case supplemented also by
generalized gradient corrections to the exchange-correlation
potential.  Plots of the spin and orbital polarization in the first
case show roughly maximal values in the localized regime, which decay
to zero in the vicinity of the transition \cite{eriksson90}.

One may also discriminate amongst the 14 $f$ orbitals purely on the
basis of their occupation, as noted in regard to the second term in
Eq.(\ref{back5}), with the size of the effect given by the monopole
integral, $F^0$, the familiar Hubbard $U_f$.  Because of the poorly
cancelled self-interaction of an electron with itself in LDF theory as
well as its spin and orbitally polarized generalizations, this effect
is largely absent from the above calculations.  It may be reintroduced
by performing self-interaction corrected (SIC) LDF calculations
generally combined with spin polarization.  Such calculations split the
occupied and empty $f$ states, determined self-consistently, by $U_f$
and therefore also serve to reduce $n(N\!-\!n)$ of Eq.~(\ref{back2})
potentially to zero, for any value of $n$, in the large volume
localized regime.  Quite satisfactory results have been obtained for
both the Ce \cite{szotek94}--\cite{svane96} and Pr \cite{svane97}
volume collapse transitions.  The LDA+U method \cite{anisimov91} is an
{\it ad hoc} way of achieving the same end in a much easier calculation
by simply adding the second term in Eq.(\ref{back5}) to the LDF-total
energy functional.  Apparently, a value of $U_f\sim 3$ eV, about half
of what would be expected, is required to make the Ce transition occur
in the right place \cite{sandalov95}.  This is interesting, since
judging from the separations between the centers of gravity of the
empty and occupied $4f$ states, SIC calculations for Ce \cite{szotek94}
and Pr \cite{temmerman93} both correspond to $U_f\sim 9$--10 eV.  It
might also be noted that $\sim$40\% smaller effective $U$ values are
also required in analytic random phase and conserving approximations
when attempting to fit exact Quantum Monte Carlo results \cite{SWB}.

\subsection{Hartree Fock with static screening}

The Hartree-Fock (HF) method has no self interaction problem.
Unfortunately, its electron-electron interactions are unscreened so
that the one-electron spectrum reflects the full value of the bare
Slater monopole integrals, e.g., $F^0\sim 26$ eV for the $4f$ states in
Ce \cite{mann67,towler94}.  If the Hamiltonian is written in second
quantized form, however, it is trivial to replace $F^0$ by a screened
value.  HF solution of such a Hamiltonian includes some correlation
effects via this static screening, and might be viewed as a further
approximation to the GW method \cite{surh98} .

We report here such Hartree-Fock calculations for the Hamiltonian,
\begin{eqnarray}
H = &&\sum_{{\bf k},\alpha,\beta} h_{\alpha\beta}^{\rm LDF}({\bf k})
\,c^\dagger_{{\bf k}\alpha}c_{{\bf k}\beta}\; +\;
\sum_{{\bf k},m,\sigma} (\varepsilon_f - \varepsilon_f^{\rm LDF})
\,c^\dagger_{{\bf k}fm\sigma}c_{{\bf k}fm\sigma}
\nonumber \\
&&+\; \frac12 U_f \sum_i \hat{n}_{if}(\hat{n}_{if}-1) \, ,
\label{mf1}
\end{eqnarray}
where ${\bf k}$ are vectors in the Brillouin zone, $\alpha\equiv
lm\sigma$ represents the usual angular, magnetic, and spin quantum
numbers, $i$ are the lattice sites, and $\hat{n}_{if} = \sum_{m\sigma}
c^\dagger_{ifm\sigma} c_{ifm\sigma}$ is the total $f$ number operator
for site $i$.  We consider only one atom per unit cell, so that the
localized HF solutions of Eq.(\ref{mf1}) will be ferromagnetic.  Rather
than the limited hopping parameter information given in
Figs.~\ref{parc}--\ref{pard}, we use the full, converged, ${\bf
k}$-dependent LDF Hamiltonians, $h^{\rm LDF}_{\alpha\beta}({\bf k})$,
which are 32$\times$32 matrices ($s$--$f$ basis plus spin) that have
been orthogonalized.  They are corrected by replacing the LDF $f$ site
energy by its constrained occupation counterpart from Fig.~{\ref{parb},
and adding the screened, $f$--$f$ monopole Coulomb interaction, though
as yet, no multipole contributions such as exchange.  Note that each of
$h^{\rm LDF}$, $\varepsilon^{\rm LDF}_f$, $\varepsilon_f$, and $U_f$ is
volume and material dependent.

Figure  \ref{cenf} summarizes the ground state HF solutions of
Eq.(\ref{mf1}) for Ce.  We have omitted the spin-orbit interaction here
for simplicity, however, we obtain similar results with this
interaction added to Eq.(\ref{mf1}).  The figure shows the total $f$
occupancy per site, $n_f\!=\!n_{if}\!\equiv\!\langle
\hat{n}_{if}\rangle$ as a function of position of the $f$ level,
$\varepsilon_f$, relative to the valence Fermi level, $\mu_v$, i.e.
that for just three $spd$ electrons per site.  Each curve corresponds
to a different volume.  The calculated values of
$\varepsilon_f\!-\!\mu_v$ from Fig.~\ref{parb} are shown as the open
circles on the appropriate curves.  The use of $\varepsilon_f$ as an
independent variable here is both instructive and offers some sense of
the impact of the $\sim 1$ eV uncertainty in these values discussed in
Sec.~3.

Note first that at large volume (solid curve,
$V\!=\!56.7\,$\AA$^3$/atom), where the hybridization interactions are
weak, one obtains a staircase structure of integer values for $n_f$.
As $\varepsilon_f$ drops below $\mu_v$, only one $f$ electron state
\begin{figure}[b]
\hspace*{-0.05in}
\vspace*{0.1in}
\psfig{file=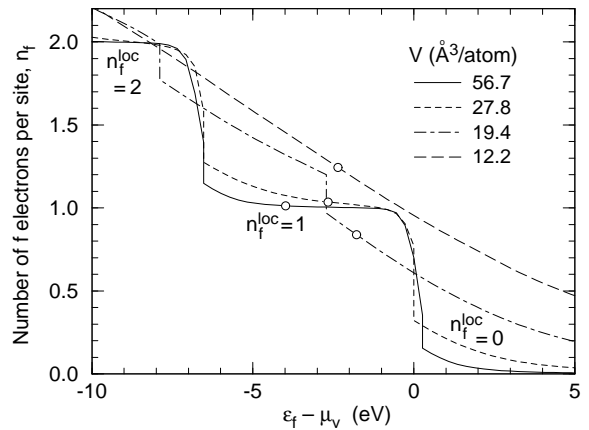,width=3.0in,angle=0}
\caption{Static-screened Hartree-Fock calculations for Ce. The number
of $f$ electrons per site, $n_f$, is shown as a function of the $f$
level position, $\varepsilon_f$, relative to the valence Fermi energy,
$\mu_v$, at four different volumes.  The calculated values of
$\varepsilon_f$ from Fig.~\protect\ref{parb} are shown by the open
circles.  The solutions exhibit different branches characterized by
integer numbers, $n_f^{\rm loc}$, of occupied split-off $f$ bands.}
\label{cenf}
\end{figure}
\noindent
is occupied at first due to the cost, $U_f$, for adding a
second.  Only when $\varepsilon_f$ is lowered an additional $U_f$ below
$\mu_v$ is the second $f$ electron picked up, and so on.  Such a plot
is similar to an integral over the $f$-decomposed density of states, so
that the plateaus correspond to Mott-like gaps in the $f$ spectrum.  A
related plot, where the chemical potential is varied, is a standard
tool used in analyzing the results of many-body simulations
\cite{plateaus}.  This staircase structure is smeared out with the
increasing hybridization width of the $f$ band as volume is reduced.
When this width becomes so large that $U_f$ is effectively unimportant,
$n_f$ grows linearly with decreasing $\varepsilon_f\!-\!\mu_v$.  The
idealized limits of localized and itinerant character are associated
with such staircase versus linear behavior, respectively.  The
usefulness of a plot such as Fig.~\ref{cenf} is that it provides a
visual sense of where a given material at a specific volume lies
between these two limits.  In particular, the short dashed curve in
Fig.~\ref{cenf} corresponds to a volume very close to that of the
collapsed $\alpha$ phase of Ce, suggesting that the Ce transition lies
closer to the localized limit.  The ratios $U_f/W_{ff}=6.7$ and
$U_f/W_{fv}=2.6$ from Figs.~\ref{parc}--\ref{pard} for Ce at the
$\alpha$ volume are consistent with this observation, and might be
compared to 3--5 and 1.1--1.5, respectively, on the low-volume side of
the other rare earth transitions.

We have labeled the HF solutions in Fig.~\ref{cenf} by integer values
of $n_f^{\rm loc}$, referring to the number of $f$ electrons split off
in low energy fully occupied bands.  The partial $f$ density of states
(DOS) shown in Fig.~\ref{cedos} illustrates the differences at
essentially the $\alpha$-Ce conditions ($V\!=\!27.8$ \AA$^3$/atom and
$\varepsilon_f\!-\!\mu_v=-2.2$ eV).  The ground state $n_f^{\rm
loc}\!=1$ solution (dashed/shaded curve) exhibits the features
associated with one localized electron: 98\% occupancy of a single
spin-orbital (total $n_f\!=\!1.05$), and a separation
\begin{figure}[b]
\hspace*{-0.05in}
\vspace*{0.1in}
\psfig{file=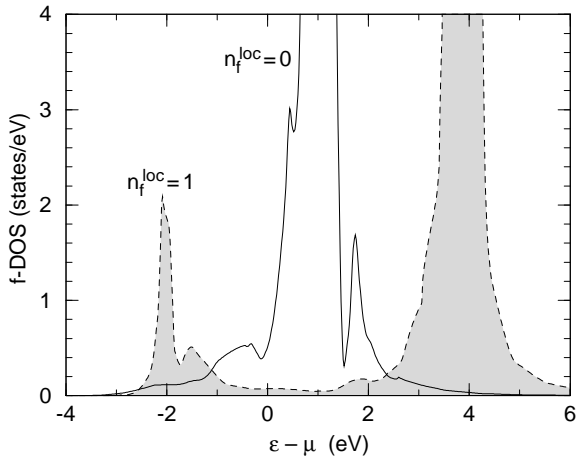,width=3.0in,angle=0}
\caption{Ce $f$ densities of states for stable $n_f^{\rm loc}\!=\!1$
and metastable $n_f^{\rm loc}\!=\!0$ Hartree-Fock solutions for Ce at
essentially the $\alpha$ phase volume and $\varepsilon_f$.  The zero of
energy is the full system Fermi energy, $\mu$.}
\label{cedos}
\end{figure}
\noindent
between occupied and empty $f$ states of $\sim U_f\!=\!6.7$ eV.  The
metastable $n_f^{\rm loc}\!=0$ solution, on the other hand, has its
$n_f\!=\!0.71$ $f$ electrons distributed across all of 14 $f$ states,
and is consistent with itinerant $f$ electron behavior.  While one
might have expected the latter solution to have the lower energy,
Sandalov et. al.  \cite{sandalov95} have found a similar result in
their tight-binding and LDA+U calculations, observing that they would
need to reduce $U_f$ to $\sim\!2.7$ eV to put the itinerant state
lower.

While it occurs at too high compression, Fig.~\ref{cenf} does
ultimately predict a transition in Ce, since the boundary between
$n_f^{\rm loc}\!=\!1$ and 0 solutions shifts to lower values of
$\varepsilon_f\!-\!\mu_v$ with compression.  The open circle for
$V\!=\!19.4$ \AA$^3$/atom (dash-dot curve) may be seen to lie on the
$n_f^{\rm loc}\!=\!0$ branch.  We find the same behavior in compressed
Pr however, the analogous transition is from $n_f^{\rm loc}\!=\!2$ to
1, as if the two $f$ electrons in Pr delocalize {\it one at a time} in
the present calculations.

The present HF results could be improved by including multipole terms
and, as noted by Sandalov et. al. \cite{sandalov95}, by using
state-dependent hopping interactions.  Nonetheless, the present results
illustrate some features which appear to be characteristic of the
 modified mean-field theories as a class.  In all cases these theories
 model the localized regime by polarized solutions with an integer
number, $n^{\rm loc}_f$, of filled split-off bands of relatively select
spin-orbital character.  The pressure induced transitions predicted by
these methods involve a shift of these bands to the vicinity of the
Fermi level, i.e., a transfer of spectral weight in roughly integer
units of electrons per atom.  Moreover, these transitions are abrupt in
the sense that there are generally two distinct solutions (e.g.,
Fig.~\ref{cedos}) with total energies which cross at some point.  To
anticipate the next section, correlation effects enable a more
continuous and nonintegral transfer of $f$ spectral weight from the
low-lying localized or polarized $f$ states to higher energies
\cite{lundin97}.  In the case of Ce, however, Svane \cite{svane96} has
claimed that such effects modify the total energy in the vicinity of
the crossing, but not further away where the Maxwell common-tangent
touches the two energy curves.

\section{EXACT QUANTUM MONTE CARLO CALCULATIONS}

As emphasized above,
electronic correlations play an important role in volume collapse
transitions, yet approximate mean-field treatments of their effects,
such as Hartree-Fock theory, suffer serious deficiencies.  We therefore
turn to an approach, Quantum Monte Carlo (QMC), which can treat
electron--electron interactions exactly.  QMC is computationally very
expensive, and while models of increasingly accurate orbital realism
can now be simulated, it is still not feasible to study a system with
the full complexity of, for example, Ce.  A crucial task is
to put together LDF calculations which do treat the full electronic
structure, with a QMC treatment of
simpler Hamiltonians which can focus on the physics
missed by LDF theory.

\subsection{Quantum Monte Carlo method}

The idea of QMC is to use an exact mathematical transformation to 
rewrite the interactions
between two electrons as a coupling of a single 
electron with a classical fluctuating field, which then mediates
the interactions indirectly.
The resulting single--particle quantum mechanical problem can be solved,
leaving only an integration over all possible values 
of the classical field.  This integration is in
an extremely high dimensional space.   For every pair of interacting
orbitals in the original Hamiltonian, there are $L$
auxiliary field variables, where $L$ 
is roughly the ratio of the bandwidth to the temperature.
Such high dimensional integrals can be done by 
classical monte carlo methods.

This formalism is straightforward and exact, but there are a number
of difficulties in its implementation.  The first is the
scaling of the computation time with system size.
While the integral to be done is classical,
the integrand is non--local.
Each field variable interacts with {\it all} the others,
not just with a few ``neighbors''.  To update every component
of the field takes a computation time which scales as the
cube of the number of orbitals.  Thus to change the number of
orbitals per site from a Hamiltonian with 
one conduction and one local orbital per site, to a fully
realistic all--orbital description in which
16 orbitals are retained, increases the 
computational expense by a factor of $8^3$.

Going to lower temperatures also involves increased effort, since the
number of field variables scales as bandwidth over temperature.
This is only a linear cost.  That is, to double $\beta=1/T$ only
doubles the cpu requirements.  However, a more serious difficulty
is also encountered.  In many situations the integrand,
which one is using as a sampling probability, can go negative.
When this occurs, QMC simulations are no longer feasible.
This ``sign problem'' does not occur for certain ``symmetric'' choices
of the parameters in the Hamiltonian, and also does not occur at 
high temperatures.  Nevertheless, it is
a significant limitation to the fillings and interaction
strengths for QMC in general.

The state--of--the--art in QMC on vector
supercomputers are simulations of Hamiltonians with 
roughly a hundred orbitals.  This means periodic lattices of, for example, 
4x4, 6x6, 8x8, and
10x10 for a two dimensional single orbital model,
and 4x4, 6x6, and 8x8 for a two dimensional two orbital
model \cite{DAGOTTO,VEKIC,MOTOME}.  Such sets
allow for extrapolations to the thermodynamic
limit, particularly when the form of the finite size
correction is known \cite{HUSE,WHITE}.  It is clear that simulations
with increased orbital realism, in three dimensions,
require significantly more powerful hardware, such as is now
becoming available with the parallel platforms.
QMC is ideally suited for parallelization; it achieves an almost linear
speedup as more and more processors are applied to a problem.  
We will discuss finite size effects in more detail later, but for
the moment let us say that we do not expect them to be too
significant for the thermodynamics, since the energy involves only
local quantities \cite{MOREO}.  Therefore our work will focus on the issue of
more orbitals per site, as opposed to increased spatial size.
This is also the direction desired for increased contact with LDF theory.

\subsection{The Hubbard and Anderson Lattice Hamiltonians}

The determinant Quantum Monte Carlo calculations we will use 
consider electrons which move on a discrete lattice
of atomic sites, and their associated orbitals \cite{BLANKENBECLER}.
Continuum QMC methods, like Green's Function Monte Carlo,
do exist, but as yet appear less easily applied to the
problems of magnetic moment formation, magnetic ordering, and 
singlet formation, which are relevant to the volume collapse transition.  
The simplest lattice model which might be applied to the present system
is the single--band ``Hubbard Hamiltonian.'' 
Originally formulated for ``d'' electron systems, 
in the present case one might consider a single effective
$f$ electron.  If we denote by $f_{i\sigma}^{\dagger}$ an
operator which creates such an electron on site $i$ with spin $\sigma$,
then this Hamiltonian may be written    
\begin{eqnarray}
H &=& -t_{ff} \sum_{\langle ij \rangle \sigma}
(f_{i\sigma}^{\dagger} f_{j\sigma}
+ f_{j\sigma}^{\dagger} f_{i\sigma}) \nonumber \\
&&+U_{f} \sum_{i} (n_{i\uparrow} - \frac12)
(n_{i\downarrow} - \frac12)
-\mu \sum_{i\sigma} n_{i\sigma}  \, .
\label{back101}
\end{eqnarray}
The kinetic energy and interactions in this model are as local as
possible, hopping between near--neighbor sites $\langle ij \rangle$
only, and interactions between the density of electrons,
$n_{i\sigma}=f_{i\sigma}^{\dagger}f_{i\sigma}$, only on the same site.
The interaction term has been written in a special form which makes
``half--filling,'' a density of one electron per site on average, occur
precisely at $\mu=0$ for any choice of temperature $T$ or parameters
$t_{ff},U_f$ in the Hamiltonian.  Furthermore there is no sign
problem.  These desirable properties require a bipartite lattice,
composed of two sublattices, such that all near neighbors of an atom on
one sublattice lie on the other.  Body-centered cubic and simple cubic
are examples, however, face-centered cubic is not.  Since, for
near-neighbor interactions, the first structure has a divergent
noninteracting density of states at $\mu=0$, the present simulations
have considered the simple cubic structure.     

The kinetic energy term in Eq.(\ref{back101}) can be equivalently
written in terms of operators $f_{k\sigma}^{\dagger}$ which create
electrons of momentum $k$ as $\sum_{k\sigma} \epsilon_{k\sigma}
f_{k\sigma}^{\dagger}f_{k\sigma}$.  For a 3--d simple cubic lattice
\begin{equation}
\epsilon_{k\sigma}= -2t_{ff}\,(\cos k_x a + \cos k_y a + \cos k_z a) \, ,
\label{back101b}
\end{equation}
where $a$ is the lattice constant.  This gives a bandwidth
$W_f=12\,t_{ff}$.  As a function of atomic volume, $V$,
Figs.~\ref{para} and \ref{parc} shows the parameters in
Eq.(\ref{back101}) to scale roughly like $U_f\sim$ constant and
$t_{f\!f}\propto W_f \sim V^{-2}$.  We shall therefore present the
phase diagram of the one band Hubbard model in dimensionless units
$T/U_f$ versus $U_f/W_f$ to approximate $T$ versus $V^2$,
respectively.

Despite its simplicity, even this single band Hubbard model cannot
be solved analytically, and while some of its properties
are fairly well established, other features, both qualitative
and quantitative, are subject to considerable debate \cite{MONTORSI}.
What is fairly certain
is that on a 3--d simple cubic lattice the ground state is an
antiferromagnetic insulator at half--filling, for all values
of the ratio $U_f/W_f$. 
The magnetic ordering can most simply be understood
as a consequence of the divergence of the noninteracting magnetic
susceptibility,
\begin{equation}
\chi_{0}({\bf q}_{*},T)={1 \over N}  \sum_{{\bf p}} { {\rm tanh} 
(\beta \epsilon_{{\bf p}} /2) \over
2 \epsilon_{{\bf p}} } \, ,
\end{equation}
at ordering wavevector ${\bf q_{*}}=(\pi,\pi,\pi)$.
Since $\chi_0({\bf q}_{*},T) \rightarrow \infty$ as 
$T \rightarrow 0$,
the Stoner criterion $U_f \chi_{0}=1$ is satisfied for any value
of the interaction strength at sufficiently low temperatures.
At strong coupling, large $U_f/W_f$, 
the insulating behavior is usually explained 
by noting that for densities less than half--filling the system
can be in configurations in which no sites are doubly occupied, that is,
no sites contain both an up and down spin electron.  
However, at densities greater than
half--filling, double occupation is inevitable, and hence the
system suddenly finds itself in a manifold of states of an energy
$U_{f}$ higher.  Thus
the density of states of the model consists
of ``upper--''
\begin{figure}[b]
\hspace*{-0.05in}
\vspace*{0.1in}
\psfig{file=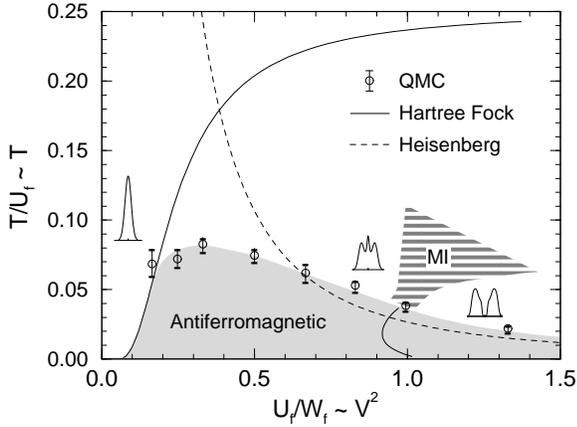,width=3.0in,angle=0}
\caption{Phase diagram of the single--band Hubbard Model; 1 effective f
orbital per site.  The N\'eel temperature $T_N$ below which
antiferromagnetic ordering takes place is indicated by the boundary
between the shaded (antiferromagnetic) and unshaded regions.  For small
$U_f/W_f$, $T_N$ is well--approximated by Hartree--Fock (solid line).
For large $U_f/W_f$, $T_N$ matches onto the spin--1/2 antiferromagnetic
Heisenberg Model (dashed line).  QMC, which is exact, picks up both
limits and the intermediate coupling regime \protect\cite{scalettar89},
which both Hartree Fock and Heisenberg fail to capture.  The inserts
show the density of states for the infinite dimensional Hubbard model,
and the continuous metal-insulator (MI) transition where the central
resonance is lost \protect\cite{jarrell93}.  The solid curve below is
an extension of this MI boundary which might apply were the magnetic
order suppressed \protect\cite{jarrell93}.}
\label{101}
\end{figure}
\noindent
and ``lower--Hubbard bands'', which are separated 
in energy by the on--site repulsion $U_f$.
If $U_{f}$ exceeds the bandwidth of the system, this gives rise to 
a ``Mott--Hubbard'' gap at half--filling.
At weak coupling, small $U_f/W_f$, these upper and lower Hubbard bands overlap,
but the system remains insulating because of the doubling
of the unit cell due to the antiferromagnetic order.

Considering now finite temperature, but remaining at half--filling, one
finds that the N\'eel temperature $T_N$ below which antiferromagnetic
ordering takes place exhibits a non--monotonic behavior with the ratio
$U_{f}/W_{f}$.  This is shown in Fig.~\ref{101}.  At weak coupling
$T_N$ grows with $U_{f}/W_{f}$ in a manner well--described by the
Stoner criterion, and manifested in the Hartree-Fock result shown by
the solid curve.  However, at strong coupling the single band Hubbard
model maps onto the quantum spin--1/2 antiferromagnetic Heisenberg
model, with exchange constant $J= 4 t_{f\!f}^2/U_f$.  Thus at strong
coupling the Hubbard model $T_N$ turns over and falls, as seen in the
QMC results (data points) in Fig.~\ref{101} \cite{scalettar89}.
Indeed, high temperature series calculations give $T_N =
3.83\,t_{f\!f}^2/U_f$ \cite{RUSHBROOK}, shown by the dashed curve in
Fig.~\ref{101}.

It is important to realize that Hartree-Fock (HF) mean field theory
completely misidentifies the strong coupling behavior of $T_N$, and,
instead of the correct Heisenberg result, predicts that $T_N$
approaches $U_f/4$ with increasing $U_f/W_f$.  At the HF N\'eel
temperature, local moments do become well--formed, but contrary to the
HF results, these moments remain unordered until much lower
temperatures set by the scale $J$. The  QMC calculations, on the other
hand, can separately resolve these two different energy scales, either
by studying the appropriate correlation functions, the local moment and
the magnetic structure factor \cite{WHITE}, or else directly from the
thermodynamics \cite{RTS3DHUB,MOREO}.

This single band model already includes a number of features of
possible interest to the volume collapse, particularly to Johansson's
Mott transition picture \cite{johansson74}.  QMC simulations find the
evolution of the density of electrons with chemical potential shows
flat plateaus indicative of a well--formed gap as the occupation per
site passes through $\langle n \rangle = 1$ \cite{WHITE}.  Unlike
mean-field treatments, however, the transfer of spectral weight as the
ratio of $U_f/W_f$ increases is much less abrupt, and includes the
development of a resonance at the Fermi surface before the formation of
upper and lower Hubbard bands, as sketched in the inserts in
Fig.~\ref{101} \cite{jarrell93}.  Calculations for the infinite
dimensional Hubbard model suggest a continuous metal-insulator (MI)
transition in the vicinity of $U_f/W_f\sim 1$ (shaded region labeled
``MI'') associated with the loss of this resonance \cite{jarrell93}.
The solid line below is an artificial extension of this boundary into
the antiferromagnetic region, which might apply were the magnetic order
suppressed, for example, by frustrating interactions.  Similar infinite
dimensional calculations for the antiferromagnetic-paramagnetic
boundary are in reasonable agreement with the three-dimensional QMC
results in Fig.~\ref{101}, suggesting the relevance of this work to the
3--d case of interest here.

The single band Hubbard model can, of course, be generalized by
including longer range hopping or Coulomb interactions.
Including multiple orbitals, however, allows one to 
describe a number of phenomena, like the screening of
f--electron spins by conduction bands, which occur in
rare earth systems like Ce.
The most commonly considered two band Hamiltonian is
the Periodic Anderson Model (PAM).
\begin{eqnarray}
H &=& \sum_{k \sigma} \epsilon_k
d_{k\sigma}^{\dagger} d_{k\sigma}
- \sum_{k \sigma} V_k
(d_{k\sigma}^{\dagger} f_{k\sigma}
+ f_{k\sigma}^{\dagger} d_{k\sigma}) \nonumber \\
&&+ U_{f} \sum_{i} (n_{if\uparrow} - \frac12)
(n_{if\downarrow} - \frac12) \nonumber \\
&&+ \sum_{i\sigma} \epsilon_{f} n_{if\sigma}
-\mu \sum_{i\sigma} (n_{if\sigma} +n_{id\sigma}) \, .
\label{back102}
\end{eqnarray}
For the simple cubic structure considered here, we have taken
\begin{eqnarray}
\epsilon_k &=& -2 t_{dd}\,[\cos{k_x a} + \cos{k_y a} + \cos{k_z a} ] \, ,
\label{back102b} \\
V_k &=& -2 t_{fd}\,[\cos{k_x a} + \cos{k_y a} + \cos{k_z a} ] \, ,
\label{back102c}
\end{eqnarray}    
where $a$ is the lattice constant.  We now have two sets of orbitals,
$d$ and $f$.  The first is ``itinerant''-- the valence or $d$ orbitals
are hybridized on neighboring sites, giving rise to a dispersive band
$\epsilon_{k}$, with no interactions, $U_d=0$.  The second is
``localized,'' with a flat, non--dispersive band $\epsilon_{f}$, and
also highly correlated, electrons of spin up and down repel with an
on--site energy $U_{f}$.  The $f$ orbitals hybridize with the itinerant
$d$ band with matrix element $V_{k}$.  There are different choices of
$V_{k}$ used in the literature, the most common being a $k$-independent
constant for which the localized and itinerant electrons hybridize on
the same site.  Our choice, Eq.(\ref{back102c}), corresponds to the
localized orbitals hybridizing with the near--neighbor itinerant
orbitals.

The case $\epsilon_{f}\!=\!0$ is termed the ``symmetric limit'' of this
model and, as in the single band Hubbard model, at chemical potential
$\mu\!=\!0$ both the localized and itinerant orbitals are precisely
half--filled for any temperature and choice of the parameters
$t_{dd},t_{fd}$, and $U_f$ for a bipartite lattice.  Again, there is no
sign problem in this limit.  Because of the particle-hole symmetric
form in which the third term of Eq.(\ref{back102}) is written, this
limit corresponds to the site energy choice
$\varepsilon_f=\epsilon_f\!-\!U_f/2= -U_f/2$ in the notation of
Fig.~\ref{parb}.  It is our intention to eventually consider both
non-symmetric limits as well as to introduce a dispersive $f$ band.
Nonetheless, the present simpler case is of some interest, especially
since Fig.~\ref{pard} suggests the $f$--valence interaction may be
the more important origin of the overall $f$-band width.  From this
perspective, the 1-band Hubbard Hamiltonian and the present Periodic
Anderson Hamiltonian may be viewed as the simplest nontrivial
correlated models within which to examine the effects of $f$--$f$ and
$f$--valence interactions, respectively.

The physics of the PAM is by no means entirely sorted out.
Again, it is at half--filling where our understanding
is most complete.  There, a competition between an ``RKKY'' energy scale 
$E_{{\rm RKKY}} \propto  J^2/W_d$
and a Kondo energy scale 
$E_{{\rm K}} \propto  W_d\,{\rm exp}(-W_d/J)$,
where $J \propto t_{fd}^2/U_f$, that determines
which of two possible ground states occurs.
If $E_{{\rm RKKY}}$ is dominant, at low temperatures
the moments on
the localized orbitals organize in an antiferromagnetic
pattern via a coupling mediated by the
itinerant band.
If $E_{{\rm K}}$ is dominant, at low temperatures
the moments on the localized orbitals form
singlets with electrons in the conduction band.
In this phase the low temperature Curie contribution
to the uniform susceptibility is suppressed, even though the
localized orbitals remain singly occupied.

The phase diagram in the $T/U_f$--$U_f/W_f$ plane for $t_{dd}=1$ eV and
$U_f=6$ eV is rather reminiscent of the corresponding phase diagram of
the single band Hubbard Hamiltonian, where in analogy to that case we
have taken $W_f\equiv 12\,t_{fd}$ for the PAM, as discussed at the end
of Section 3.  As shown in Fig.~\ref{102}, there is a low temperature
antiferromagnetic phase, whose transition temperature exhibits a
non--monotonic dependence on $U_f/W_f$ similar to the one band case.
There is also a cross--over temperature which separates the
paramagnetic phase into a
\begin{figure}[b]
\hspace*{-0.05in}
\vspace*{0.1in}
\psfig{file=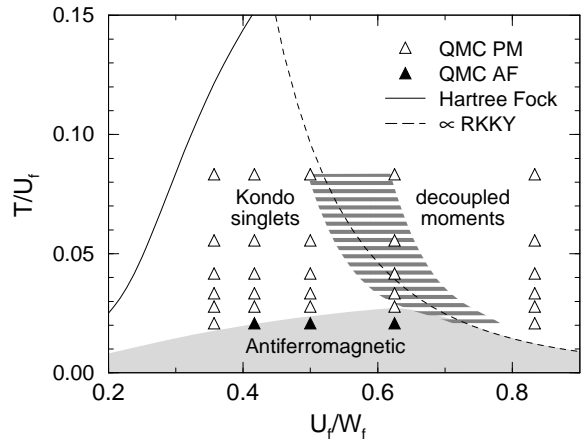,width=3.0in,angle=0}
\caption{Phase diagram of the 3--d, symmetric Periodic Anderson Model
(PAM) at half--filling (1 electron per orbital per site) from QMC
simulations.  Here $U_f = 6 t_{dd}$ is fixed, $W_f=12\,t_{fd}$ is
varied, and the number of sites $N = 4 \times 4 \times 4 = 64.$ QMC
data points are shown by triangles, the dark triangles indicate
antiferromagnetic ordering in the f--band.  The antiferromagnetic
region is indicated by shading.  At higher temperatures and high
$U_f/W_f$ free f--band moments still exist, but are decoupled and not
ordered.  As $U_f/W_f$ decreases, the f--band electrons begin to form
singlets with the conduction--band electrons, until in the regime
labeled ``Kondo singlets'' the Kondo--like singlets are fully--formed.
The hashed region marks the Kondo singlet formation cross--over.
Experimentally the two regimes are distinguished by the absence of a
Curie--like $1/T$ divergence of the uniform susceptibility in the Kondo
regime.  Hartree Fock and RKKY curves are shown for comparison.}
\label{102}
\end{figure}
\noindent
region with decoupled moments (to the right of the hatched boundary),
and a region where these moments form singlets with electrons in the
conduction band (to the left).  This cross--over region is the
counterpart of the metal-insulator transition in Fig.~\ref{101}, in
that in both cases this boundary is associated with appearance of the
central resonance for decreasing $U_f/W_f$.  The Hartree-Fock
prediction for $T_N$ (solid curve) in Fig.~\ref{102} also has the same
appearance as in Fig.~\ref{101}, first rising and then saturating at
$U_f/4$ (not shown) for increasing $U_f/W_f$.  The RKKY energy, $E_{\rm
RKKY}/U_f \propto (U_f/W_f)^{-4}$ for constant $W_d$ and $U_f$, is
analogous to the Heisenberg curve in Fig.~\ref{101}.  The ratio $E_{\rm
RKKY}/U_f$ is plotted as the dashed curve in Fig.~\ref{102}, and has
been scaled  to be consistent with the QMC location of $T_N$ near
$U_f/W_f=0.6$.

The parameters in Figs.~\ref{para}--\ref{pard} may be used to translate
the location of the observed rare earth volume collapse transitions (or
symmetry changes in Nd and Sm) into values of $U_f/W_f$.  Considering
only the $f$-valence hopping, these are in the range $U_f/W_f=$
1.1--1.9 except for Ce which is 3.7--2.6, as noted earlier.  It is not
unreasonable to associate these values with the cross--over region just
discussed, $U_f/W_f\sim 0.8$ or larger if extrapolated to room
temperature, whereas by contrast, the Hartree-Fock predictions for the
symmetric two-band model indicate a transition for $U_f/W_f\sim 0.2$.
It should also be emphasized that the one and two-band models discussed
here correspond to half-filled $f$ bands, and that lower $f$-band
filling should favor the ``itinerant'' states, pushing transitions to
larger values of $U_f/W_f$, as may occur with Ce.

%
%
%

\subsection{Density of States in the PAM}

The behavior of the quasiparticle density of states (DOS)
$N(\omega/U_f)$ provides an especially graphic illustration of how QMC
differs from mean field theories such as HF in its treatment of
correlations.

In Fig.~\ref{108} we show 
$N(\omega/U_f)$ of the localized band of the PAM for different 
inverse hybridizations $U_f/W_f$ at a temperature above the N\'eel temperature
$T_N$ at which antiferromagnetic ordering occurs.
The DOS was obtained by numerical analytic continuation of the single--particle
imaginary--time Green's function, 
$G(\tau) = \frac{1}{N}\sum_{\bf p} \langle T_{\tau} c_{\bf p}(\tau) 
c_{\bf p}^{\dagger}(0) \rangle$, using
the relation
\begin{eqnarray}
G(\tau) = \int_{-\infty}^{\infty} d (\omega/U_f) {e^{-\tau \omega/U_f}
\over  {1 + e^{-\omega / T}}} N(\omega/U_f)
\end{eqnarray}
\begin{figure}[b]
\hspace*{-0.05in}
\vspace*{0.1in}
\psfig{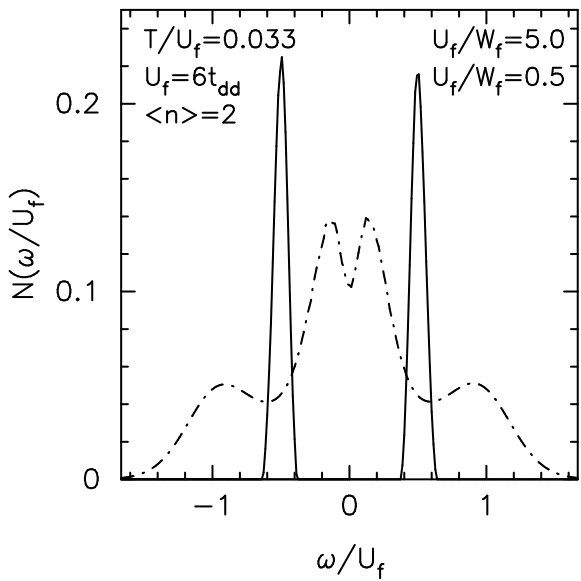}
\caption{The f-band quasi-particle density of states $N(\omega)$ as a
function of distance from the Fermi energy ($\omega_F = 0$) of the PAM
for $T/U_f = 1/30$, $U_f = 6 t_{dd}$, and at half-filling $\langle n
\rangle = 2$.  Two values of $U_f/W_f$ are shown.  For $U_f/W_f$ large, the
two Hubbard-band peaks contain all of the weight in the DOS (solid
line).  [The solid line values have been reduced by a factor of $3$.]
As $U_f/W_f$ decreases, the Kondo-singlet formation regime is entered and
spectral weight shifts continually to the center, Fermi--energy
region.  The dot--dashed line represents a small $U_f/W_f$ where the
singlets are well--formed.  The pseudo--gap, or reduction in the
spectral function $N(\omega)$ at the Fermi energy, is due to the
insulating nature of the symmetric, half--filled PAM at small $U_f/W_f$.}
\label{108}
\end{figure}
\noindent
for imaginary time $\tau$, temperature $T/U_f$, and frequency $\omega/U_f$.
We use the maximum entropy method \cite{GUBERNATISandJARRELL}
to perform the analytic continuation
in a technique which utilizes the full imaginary--time covariance matrix.

For a large $U_f/W_f$ (solid line), the spectral weight is concentrated
in the lower and upper 'Hubbard' bands, corresponding to
singly--occupied localized orbitals and to doubly--occupied localized
orbitals, respectively.  As $U_f/W_f$ is decreased, the hybridization
between the bands increases and the spectral weight simultaneously
begins to shift continuously from the two peaks to the region near the
Fermi energy ($\omega/U_f=0$).  The build--up of the central peak in
the DOS corresponds to the formation of singlets between electrons on
the localized and itinerant orbitals as will be discussed shortly.  At
a sufficiently small $U_f/W_f$ a fully--developed three--peak structure
is seen in the DOS (dot--dashed line).  The pseudo--gap, or reduction
in the spectral function $N(\omega)$ at the Fermi energy, is due to the
insulating nature of the symmetric, half--filled PAM at small $U_f/W$.

There is no mean--field analogy either for the three peak structure in
the DOS, or for this continuous shift in the spectral weight.  At these
same conditions, HF yields a partial $f$-DOS similar to Fig.~\ref{108}
for the localized limit, $U_f/W_f=5$.  However, by $U_f/W_f=0.5$, these
same lower and upper Hubbard peaks are still present, although
considerably broadened.  Not until $U_f/W_f=0.22$, at $T/U_f=0.033$, do
the HF antiferromagnetic and paramagnetic free energies cross, at which
point a single broad central HF $f$-DOS peak is observed for smaller
$U_f/W_f$.

\subsection{Magnetic Properties of the Periodic Anderson Hamiltonian}

The ultimate goal of using QMC to compute precisely the
contribution of the correlation energy for
the volume collapse transition is still in the future,
both in determining the specific appropriate lattice Hamiltonian
and in then carrying out the computations.
Here we will show some further results which illustrate
that QMC can sensitively pick up the basic physics
of correlations between moments like Kondo singlet formation and
long range antiferromagnetic order.
We will continue to simulate the PAM, Eq.(\ref{back102})
for $t_{dd}=1$ eV, $U_f=6$ eV, and a range of $t_{fd}$ values indicated as
before by $W_f=12\,t_{fd}$.   These results will provide some of the
background to the phase diagram shown in Fig.~\ref{102}.                        

While we are primarily interested in determining the thermodynamics
to infer the equation of state, it
is useful to look at magnetic properties
such as the local moment and magnetic structure factor.  In Fig.~\ref{103}
we show the square of the local moment 
$\mu_{d(f)}^2=\langle {\bf S}_{id(f)} \cdot {\bf S}_{id(f)} \rangle$
on the itinerant ($d$)
and localized ($f$) orbitals.  $\mu_d^2$ takes on a

\begin{figure}[b]
\hspace*{-0.05in}
\vspace*{0.1in}
\psfig{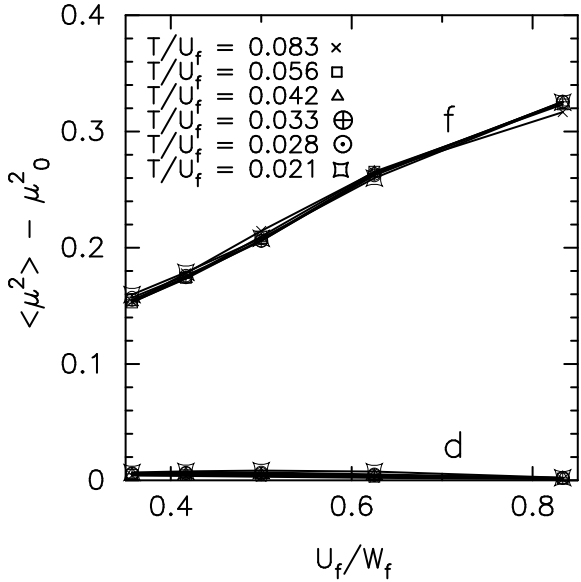}
\caption{Square of the local moment on the itinerant (lower curve) and
localized (upper curve) orbitals at various temperatures $T/U_f$
and $U_f/W_f$.  The values shown have the total moment for a non-interacting
system subtracted, so that $0$ on the curves corresponds to the moment of
a non--interacting electron system.  The itinerant orbital electrons
remain at approximately the non-interacting moment value.
The local (f) orbitals have increasingly well--formed moments
as $U_f/W_f$ becomes larger.}
\label{103}
\end{figure}
\noindent
value
essentially equal to
that for a non-interacting electron gas.
This is a consequence of the lack of correlations,
$U_d=0$, between up and down spin electrons.
In contrast, $\mu_f^2$ shows a significant 
enhancement throughout the range of interband hybridization $t_{fd}$.

$\mu_f^2$ is the most local measure possible of 
magnetism-- the presence or absence of a moment on a site.
The magnetic structure factor determines the degree of the order
between spins on different sites across the entire lattice,
\begin{equation}
S_{ff}(T,{\bf q}_{*}=(\pi,\pi,\pi)) = {1 \over N} \sum_{ij} (-1)^{i+j}
{\bf S}_{if} \cdot {\bf S}_{jf} \,.
\end{equation}
This quantity is shown in Fig.~\ref{104} for various values of the
temperature and interorbital hybridization, $t_{fd}$,
reflected in the ratio $U_f/W_f$.
A careful finite size scaling study is necessary before 
rigorous conclusions concerning the presence of long range 
antiferromagnetic order are drawn.  
The above data, however, suggest a strong tendency for the moments
to order antiferromagnetically, with an ordering temperature which
is maximal around $U_f/W_f \approx 0.6$.

As discussed above,  singlet formation between 
the spins of the localized and 
itinerant electrons competes with antiferromagnetic order.
Fig.~\ref{105} shows a quantity which measures this tendency.
Here, $c_{fd}$ is essentially $\langle {\bf S}_{if}\cdot
{\bf S}_{jd}\cdot \rangle$ averaged over near-neighbor pairs, $i,j$,
and 
\begin{figure}[b]
\hspace*{-0.05in}
\vspace*{0.1in}
\psfig{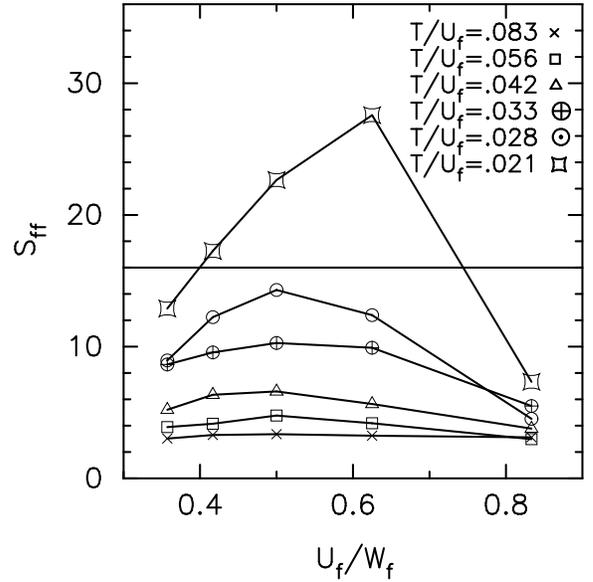}
\caption{Anti-ferromagnetic structure factor, \mbox{$S_{ff}(T/U_f,{\bf
q}_{*}=(\pi,\pi,\pi)\,)$} for various values of $U_f/W_f$.  The
horizontal line demarks the existence of antiferromagnetic long--range
order in the system.}
\label{104}
\end{figure}
\noindent
appropriately normalized to reflect the number of such pairs and
the average on-site local moments.
The crossing of these curves through
40--60 percent of their lowest saturated
value, indicated
by the horizontal lines,
gives an estimate of the Kondo temperature below which singlet
formation takes place \cite{GETTK}.

From the results shown in Figs.~\ref{103}--\ref{105},
and analysis of other quantities such as the
density of states, we infer the
\begin{figure}[b]
\hspace*{-0.05in}
\vspace*{0.1in}
\psfig{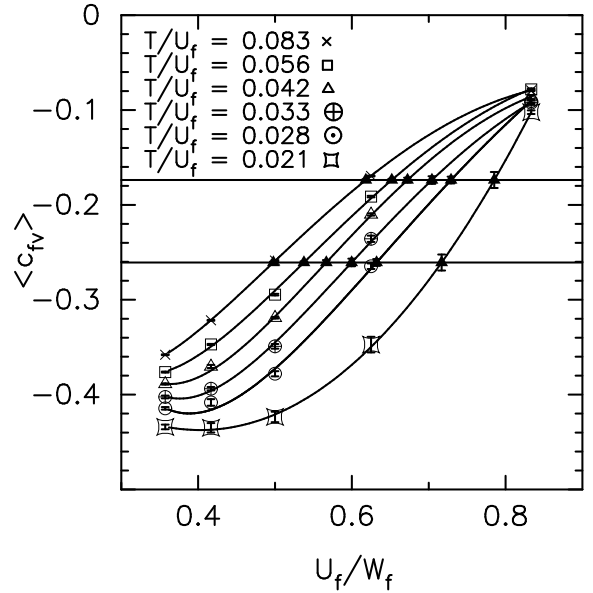}
\caption{A measure of the tendency for Kondo singlet formation.  The
horizontal lines indicate the approximate range of the cross--over
between the paramagnetic and Kondo singlet phases (hatched region of
Fig.~\protect\ref{102}.)}
\label{105}
\end{figure}
\noindent
magnetic phase diagram in the $(U_f/W_f$, $T/U_f)$ plane for
the PAM on a 3--d simple cubic lattice at $t_{dd}=1$ eV,
$U_f=6$ eV, and half--filling, shown in Fig.~\ref{102}.

Low--temperature HF calculations of the squared local moments,
$\mu^2_{d(f)}$, resemble the QMC results in Fig.~\ref{103} for the
antiferromagnetic phase.  However, {\it neither} $f$ or $d$ moment
shows any enhancement in the paramagnetic
phase, demonstrating the well
known fact that moment formation and magnetic order coincide in HF.
Both antiferromagnetic and paramagnetic solutions exhibit values of
$\langle {\bf S}_{1f}\cdot {\bf S}_{2d}\rangle$ which decrease from
$\sim 0$ to $\sim -1/4$ as $U_f/W_f$ is reduced, where 1 and 2 are the
two sites of a doubled unit cell.  While a crude immitation of the
Kondo-like behavior in Fig.~\ref{105}, even at $T\!=\!0$, this
variation is quite gradual, and the mean--field expectation can of
course never approach the full singlet limit of $-3/4$.      

\subsection{Thermodynamic Properties of the Periodic Anderson Hamiltonian}

The energy can be easily measured as a function of temperature
in QMC.  From this data, the specific heat can be obtained, for example
using a finite difference approximation for the 
derivative at two relatively closely spaced temperature values.
A more useful way to proceed is to measure $E(T)$ on a sufficiently
fine grid of temperatures and 
fit to a physically motivated functional form.
We have chosen
\begin{equation}
E(T)=E(0)+\sum_{n=1}^{N} c_{n} e^{-\beta n \Delta} \, ,
\end{equation}
where $\Delta$ and $c_n$ are free parameters.  $N$ is chosen to be 5--6
to compromise between allowing sufficient flexibility to pick up the
full structure without overfitting.  A check on this procedure is
provided by computing the integrated area,
\begin{equation}
\int_{0}^{\infty} dT {C(T) \over T} =\sum_{n=1}^{N} {c_n \over n \Delta}
=  4 \, {\rm ln}\,  2 -S_0 \, .
\label{sumrule}
\end{equation}
Here the presence of the term $S_0$ is a result of the use of a finite
periodic cluster where the entropy does not go to zero at $T=0$
\cite{ENTROPY}.  We find this sum rule on the total entropy is obeyed
to within less than 2 percent for $t_{fd}\geq 0.8$.  For smaller
$t_{fd}$ the system orders antiferromagnetically at temperatures below
those accessed in the simulations ($T \geq 0.1$ eV), and the
integral instead approaches $3\, {\rm ln}\, 2 - S_{0}$, correctly
reflecting the missing entropy of spin ordering.  Note that the QMC
calculations were extended to sufficiently high temperatures to
establish the large $T$ asymptotic dependence of $C(T)/T$.  We have
also tried other fitting forms, and find our results for the specific
heat are rather insensitive to the particular functions involved.

\begin{figure}[b]
\hspace*{-0.05in}
\vspace*{0.1in}
\psfig{file=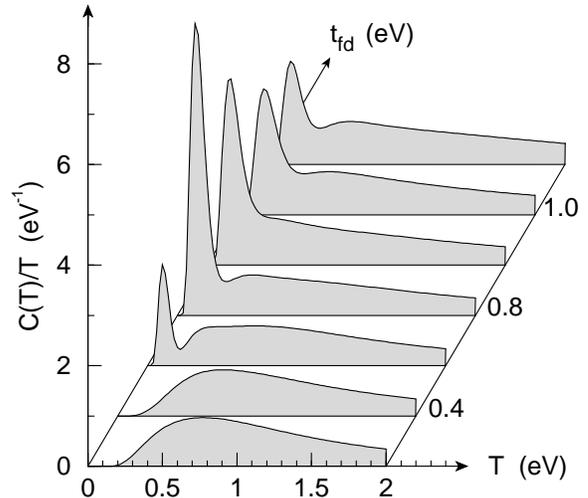,width=3.0in,angle=0}
\caption{Specific heat of the PAM obtained by fitting the QMC
energies.  The area under the curves satisfy the sum rule,
Eq.~\ref{sumrule}.  The peaks in the specific heat appear to be
associated with antiferromagnetic ordering.  (See, also,
Fig.~\protect\ref{102}.) The corresponding peaks at $t_{fd}=0.2$ and
0.4 lie below $T<0.1$ eV, the temperature range sampled by QMC, and so
are absent.}
\label{106}
\end{figure}
\noindent

The results of this procedure are shown in Fig.~\ref{106}.
The specific heat exhibits a broad bump at high temperatures,
the energy scale $U_f$ of charge fluctuations,
that is, excitations 
to the manifold of doubly occupied states.
At low temperatures there is a much sharper peak which,
given the data on the magnetic correlations,
we believe is associated with antiferromagnetic ordering.
We do not see any sharp feature associated with singlet formation,
which drives the volume collapse transition according to the 
Kondo picture \cite{allen82,lavagna82}.
However, it is well known that the signature of singlet 
formation in the specific heat becomes sharp only 
for relatively small $t_{fd}$, where, in this symmetric 
half--filled model, antiferromagnetism strongly 
competes \cite{KSHARP}.

The resolution of this difficulty is to study the PAM in parameter
regimes where the antiferromagnetic order is suppressed, either by
going away from the symmetric limit $\epsilon_f=0$ or else away from
half--filling $\mu=0$.  As discussed above, doing so is also suggested
by the LDF analysis, where the $f$ site energy, $\varepsilon_f =
\epsilon_f-U_f/2$, is seen in Fig.~\ref{parb} to rise with compression
relative to the valence chemical potential.  Moreover, HF calculations
show there is a significant suppression of the N\'eel temperature by
changing $\epsilon_f$ or $\mu$.  We have found this result using the
difference in HF energies between the antiferromagnetic and
ferromagnetic solutions as an estimate of $T_N$.  This approximation
gives the correct strong coupling Heisenberg scaling $T_N \propto
t^2/U$ in the single band model, as opposed to $T_N \propto U$ which is
the formally correct HF prediction obtained from looking at the
temperature at which the free energy of the antiferromagnetic solution
becomes lower than the paramagnetic one.

HF calculations of $C(T)/T$ for the same $4^3$-site two-band model bear
only a gross resemblance to Fig.~\ref{106}.  There is the appearance of
a magnetic ordering peak caused by the drop in $C(T)/T$ from
antiferromagnetic to paramagnetic solutions with increasting $T$ at
$T_N$.  However, this occurs at the HF $T_N$ which incorrectly
approaches $U_f/4=1.5$ eV, rather than $0$, as $t_{fd}\rightarrow 0$.
Assuming the same band widths, $W_{fd}$, the rare earth transitions
would occur for values of $t_{fd}$ in the range 0.1--0.3 eV for
near-neighbor interactions in the simple cubic structure.  At
$t_{fd}=0.2$ eV, the HF paramagnetic $C(T)/T$ is in reasonable
agreement with the corresponding curve in Fig.~\ref{106} except for
having a larger area ($4\ln 2\!-\!2S_0$ versus $3\ln 2\!-\!S_0$).  In
the experimentally observed paramagnetic regime, the HF paramagnetic
calculation provides a better approximation to the true thermodynamics
than does the HF antiferromangnetic solution even though the latter may
have lower free energy.

\section{SUMMARY}

This paper has discussed our current experimental and theoretical
understanding of the volume collapse transitions in the trivalent rare
earth metals.  The data exhibit ``localized'' $4f$-electron behavior at
low pressure, characterized by isolated-ion-like magnetic moments and
the absence of these electrons in the crystal bonding, and
``itinerant'' $4f$-electron behavior at higher pressures, without
apparent moments but with clear evidence of $4f$ bonding
participation.  The two regimes are separated by first order phase
transitions which in three cases (Ce, Pr, and Gd) are accompanied by
unusually large (9\%--15\%) volume changes.  It is generally believed
that these volume collapse transitions are associated with change in
the degree of $4f$ electron correlation.  Consistent with this idea, we
have reported calculations of relevant parameters which indicate the
transitions occur in the range $U_f/W_f = 1$--2,  where $U_f$ is the
onsite Coulomb repulsion between $4f$ electrons, and $W_f$ is a measure
of the $4f$ band width including both $f$--$f$ and $f$--valence
hybridization effects.  The one exception is Ce, where the transition
occurs at a larger ($\sim 3$) value of this ratio which may reflect the
low ($f^1$) $f$-band filling in this case.

We have compared a variety of orbitally realistic, mean-field based
methods used to calculate these transitions: spin and orbital polarized
local density functional theory, self-interaction corrections, the
LDA+U method, and Hartree-Fock (HF) with static screening.  As a class,
these methods represent the localized phases by polarized solutions
with an integer number of occupied, split-off $4f$ bands, and the
itinerant phases, by a single broad collection of $4f$ bands
overlapping the Fermi level.  The transition is therefore accompanied
by an abrupt transfer of spectral weight (the split-off bands) to the
vicinity of the Fermi level.  Both the loss of moment and the onset of
$4f$ bonding are clear from these calculations.  At very small and
large volumes, $U_f<<W_f$ and $U_f>>W_f$, respectively, such mean-field
solutions should provide the correct ground states, aside from
mistreatment of intraatomic correlation for multi-$f$ electron atoms in
the latter limit.  Indeed, Anderson has used the HF solution to
elucidate the limits of the one-band Hubbard model \cite{anderson71}.
The problem lies in between these limits where competition between the
two energy scales leads in general to a correlated solution.  Thus, while
several calculations have reported reasonable agreement with experiment
for the location of the transitions \cite{eriksson90}--\cite{svane96},
there are substantial differences between HF and exact Quantum Monte
Carlo (QMC) solutions for few band Hamiltonians which raise doubts
about these approaches.

To assess correlation effects, we have reviewed exact QMC solutions for
the one band Hubbard model, and reported new QMC results for the
two-band periodic Anderson lattice model.  We contrast both solutions
with HF results for the same Hamitonians.  The QMC results for both one
and two-band models show non-monotonic behavior of the N\'eel
temperature, $T_N$, as a function of $U_f/W_f$, where this ratio scales
like volume to a power $\sim 2$.  $T_N$ first rises with increasing
$U_f/W_f$ as antiferromagnetic order accompanies moment formation, and
then falls as decreasing exchange interactions lead to moment
disorder.  HF misses the latter effect, so that its $T_N$ continues to
rise to a large saturation value of $U_f/4\sim 1.5$ eV.  The HF
transitions are therefore always intrinsically associated with both
moment formation and magnetic ordering.  The region of magnetic order
is greatly suppressed in the QMC solutions, e.g., lying below $\sim
0.1$ eV for the present PAM results based on realistic parameters.
This value can be further suppressed by changing the $f$ level position
(non-symmetric PAM), consistent with the experimental observation that
the volume collapse transitions occur entirely within the paramagnetic
regime.

The dramatic new feature in the correlated QMC solutions, which has no
counterpart in mean field, is the three peak structure in the $f$
density of states (DOS) found especially at temperatures just above the
region of magnetic order.  The growth of the central Fermi-level
resonance for decreasing $U_f/W_f$ comes from a {\it continuous}
transfer of spectral weight from the outer split peaks, and is
accompanied in the QMC results with reduction of the magnetic moment as
might be deduced from the magnetic susceptibility.  This reduction
occurs at considerably larger values of $U_f/W_f$ or volume than does
the HF transition, and is in much better agreement with our
``experimental'' $U_f/W_f$ placement of the collapse transitions.

The central resonance is the well known signature of the Kondo effect
for the PAM, and in its impurity Anderson model context, the critical
aspect of the Kondo volume collapse model for Ce
\cite{allen82,lavagna82}.  However, it is intriguing that the infinite
dimensional one-band Hubbard model reduces formally to the impurity
Anderson model and exhibits a similar central resonance, which might be
associated with $f$ electrons screening their own moments
\cite{jarrell93}.  This blurs the distinction between the Mott
transition picture \cite{johansson74}, traditionally associated with
the Hubbard model, and the Kondo volume collapse
\cite{allen82,lavagna82}.  The distinction is further complicated in
any mean field approach which would pick up only half (e.g.,
$f_\uparrow d_\downarrow$ but not $f_\downarrow d_\uparrow$) of a Kondo
like singlet, and might therefore still look like $f$ magnetic order.
This is not the case for QMC calculations which can separately identify
reduction in the local moments as well as their screening by the
valence electrons, as seen in the present PAM results.  More definitive
use of QMC in this manner, however, awaits studies in other parameter
regimes (e.g., non symmetric PAM) in search of thermodynamic features,
for example, which might unambiguously demonstrate that the given model
does indeed capture the essential nature of the rare earth volume
collapse transitions.

A natural future direction for the theory of the rare earth volume
collapse transitions would be to compromise between the all-orbital
mean-field based methods and few-orbital exact treatments, so as to
incorporate only the most important correlation effects within a
sufficiently rich multi-orbital framework.  There appears to be, for
example, a growing awareness that some form of dynamic screening is
critical to incorporate features such as the three peak density of
states, requiring inclusion of an energy-dependent self-energy.  Some
success has been achieved with local, {\bf k}-independent
approximations to such self energies
\cite{steiner92,han97,lichtenstein98}.  In principal this approach
reduces to a correlated impurity problem which could be solved by QMC
methods for far more orbitals per site that possible in the fully
correlated lattice.  Should intermediate or longer-range correlations
be critical, a new modified dynamical mean field technique may allow
inclusion of such correlations in addition to multiple bands
\cite{DCA}.  All of these techniques form a bridge between local
density functional methods which serve to define realistic
multi-orbital Hamiltonians which may then be solved by correlated
approaches such as QMC.

\section*{Acknowledgments}

Work at LLNL was performed under the auspices of the U.S. Department of
Energy by Lawrence Livermore National Laboratory under Contract No.
W-7405-Eng-48.  The authors at U.C. Davis would like to acknowledge
support from the Associated Western Universities and from the LLNL
Materials Research Institute.



\vspace{3in}
\noindent
Prepared for a 1998 issue of the {\it Journal of Computer-Aided
Materials Design}


\begin{thebibliography}{99}
\bibitem[86]{ENTROPY}  The value of $S_0$ is determined by the
one--body part of the Hamiltonian.  It has been separately calculated
and agrees with the QMC results.  $S_0=0.4332$ for Eq.(\ref{back102b})
and $4^3$ sites.
\end{thebibliography}
\end{document}